\documentclass[a4paper]{article}

\usepackage[T1]{fontenc}
\usepackage[utf8]{inputenc}
\usepackage{amsmath,amssymb,slashed}
\usepackage{amsfonts}
\usepackage{float}
\usepackage{upgreek}
\usepackage{comment}
\usepackage{graphicx}
\usepackage[mathscr]{eucal}
\usepackage{epsfig}
\usepackage{booktabs}
\usepackage{bbold}
\usepackage{slashed}
\usepackage[numbers,sort&compress]{natbib}
\usepackage[colorlinks,linkcolor=red,anchorcolor=green,citecolor=blue]{hyperref}

\numberwithin{equation}{section}

\begin{document}

	\thispagestyle{empty}
	
	\begin{flushright}\footnotesize
		\vspace{2 cm}
	\end{flushright}

	\begin{center}
		{\Large\textbf{\mathversion{bold} Theory of packaged entangled states}}
		
		\vspace{1cm}
		
		\textrm{Rongchao Ma*}
		
		\textit{Technical Institute of Physics and Chemistry, Chinese Academy of Sciences, Beijing 100190, China}
		
		\vspace{1cm}
		
		\textrm{*Corresponding author. \\ E-mail address: marongchao@yahoo.com (Rongchao Ma)}
	\end{center}

\begin{abstract}
The entangled states that include every physical properties of particles would be important for both theoretical and applied physics. However, the existence and properties of such entangled states are unclear at present. 
Here we theoretically show that a particle-antiparticle pair can form the so-called packaged entangled states which encapsulate all the necessary physical quantities for completely identifying the particles. The particles in the packaged entangled states are indeterminate and exhibit unusual properties.
Thereafter, we discussed the possible applications of these new entangled states, i.e., the protocol for teleporting the entire quantum state of a particle (or an antiparticle) to an arbitrarily large distance without a classical channel, transfer of the new entangled states from a particle pair to another particle pair, and new interpretation to the matter-antimatter asymmetry of the observable universe. 
\end{abstract}

\section{Introduction}

An entangled state usually refers to a pure state of a composite system which cannot be expressed as the direct product of the quantum states of its subsystems.\cite{Horodecki,Lo,Nielsen} In these states, a measurement on one of the particles will immediately change the state of other particles \cite{Einstein} via the ``spooky action at a distance'' \cite{Einstein2} no matter how far these particles are spatially separated. The entangled states are important for both fundamental research \cite{Einstein,Bell,Freedman,Aspect} and applications in quantum information \cite{Bennett}. 
A considerable amount of theoretical and technical works have been devoted to the entanglement of one degree of freedom \cite{Bouwmeester,Boschi,Leuenberger,Pfaff,Krauter,Hofmann,Riebe,Barrett,Olmschenk,Nakamura,Baur} and several degrees of freedom, i.e., hyperentanglement \cite{Kwiat,Barreiro,Chen,Vallone,Liu}, multimode entanglement \cite{Gatti,Giovannetti,Tan,Shi,Liew,Knott}, and entanglement of indistinguishable particles  \cite{Kulik,Marinatto,Marzolino}. These entangled states usually encapsulate part of the particles' physical quantities.

However, one may ask whether the particles can form a special entangled state that can package all their physical properties?  
More specifically, a particle possesses a number of independent physical quantities (or freedoms), such as charge, baryon number, lepton number etc.\cite{Griffiths,Perkins} With these physical quantities, one can completely identify the particles. The question is whether all these physical quantities could be packaged as an entirety in the new entangled state. 
We shall call such an entangled state as a packaged entangled state, which is different to the states of hyperentanglement \cite{Kwiat,Barreiro,Chen,Vallone,Liu} or multimode entanglement \cite{Gatti,Giovannetti,Tan,Shi,Liew,Knott}.

The packaged entangled states could be used to teleport the entire quantum states or to transmit every intrinsic property of the particles or antiparticles \cite{Krauss}, instead of just some of their physical properties. On the other hand, the packaged entangled states could be also used to explain the imbalance between baryons and antibaryons produced after the Big Bang.\cite{Dine} Finally, the packaged entangled states could be also applied in medicine and energy transportation.

This paper is organized as follows. 
In section \ref{Theo}, we first constructed the mathematical expressions of the packaged entangled states of a particle-antiparticle pair and discussed their properties. 
Next, we discussed how to teleport a particle/antiparticle to a place at a distance using the packaged entangled states. The advantage of this protocol is that the classical channel is removed due to the particle-antiparticle annihilation phenomenon. Thirdly, we discussed how to transfer a packaged entangled state from two particles to two other particles. Finally, we showed that the collapse of packaged entangled states contributes to the origin of matter-antimatter asymmetry (or baryogenesis) of the observable universe.
In section \ref{Disc}, we compared the difference between the packaged entangled states and the entangled states studied in early literatures.

\section{Theory}
\label{Theo}

Because the packaged entangled states encapsulate all the necessary physical quantities for completely identifying the particles, they should not include some particular physical quantity symbols but include the abstract particle symbols, i.e., the particle quantum state $\left|P\right\rangle$ and antiparticle quantum state $\left|\bar{P}\right\rangle$. We further require the packaged entangled states to be the eigenstates of some operator for the possible experimental test of their existence. Let us now use charge conjugation to construct the mathematical expression of the packaged entangled states of a particle-antiparticle pair.

It is known that $\left|P\right\rangle$ and $\left|\bar{P}\right\rangle$ are symmetrical in the sense of charge, i.e., the charge of $\left|P\right\rangle$ and $\left|\bar{P}\right\rangle$ are equal in quantity but with opposite signs. From particle physics \cite{Griffiths,Perkins,Peskin} we know that $\left|P\right\rangle $ and $\left| \bar{P} \right\rangle$ can be interchanged by the charge conjugation operator $C$, i.e., $C \left|P\right\rangle = \left|\bar{P}\right\rangle$ and $C \left|\bar{P}\right\rangle = \left|P\right\rangle$. This mutual transformation indicates that $\left|P\right\rangle $ and $\left| \bar{P} \right\rangle$ could be superimposed in the sense of wave function under certain conditions. In fact, the superposition states with different charge exist in the systems like the well-studied bound ``particle-antiparticle'' pairs  $\left|P\bar{P}\right\rangle$, which are the eigenstates of $C$.\cite{BoundPair} One can also find the similar discussions about the superposition states with different charge in early literatures \cite{Aharonov,Rolnick}.

The charge conjugation not only reverses the sign of the particle's electric charge ($Q$), but also reverses the sign of all other internal quantum numbers \cite{Griffiths}, i.e., baryon number ($B$), lepton number ($L$), isospin ($I_3$), charm ($C$), strangeness ($S$), topness ($T$), and bottomness ($B'$). All these internal quantum numbers are packaged together under the charge conjugation. However, the charge conjugation does not change the particles' mass, energy, momentum, and spin.\cite{Griffiths}

\subsection{Packaged entangled states with C-symmetry}

We first wish to construct the packaged entangled states with C-symmetry. Let us now consider the following separable states of a particle-antiparticle pair $A$ and $B$, i.e., 
\begin{subequations}
	\label{ThetaS}
	\begin{eqnarray}
	\left| \Theta^+ \right\rangle_{AB} &=& \left| P \right\rangle_A \left| \bar{P} \right\rangle_B, \\
	\left| \Theta^- \right\rangle_{AB} &=& \left| \bar{P} \right\rangle_A \left| P \right\rangle_B,	
	\end{eqnarray}
\end{subequations}
and their linear combinations (superpositions), 
\begin{subequations}
	\label{PsiE}
	\begin{eqnarray}	
	\left| \Psi^+ \right\rangle_{AB} &=& \frac{1}{\sqrt{2}} \left(\left| P \right\rangle_A \left| \bar{P} \right\rangle_B + \left| \bar{P} \right\rangle_A \left| P \right\rangle_B  \right), \label{PsiP}  \\
	\left| \Psi^- \right\rangle_{AB} &=& \frac{1}{\sqrt{2}} \left(\left| P \right\rangle_A \left| \bar{P} \right\rangle_B - \left| \bar{P} \right\rangle_A \left| P \right\rangle_B  \right). \label{PsiM}
	\end{eqnarray}
\end{subequations}

Apply the charge conjugation operator $C$ to the separable states, we have $C\left| \Theta^+ \right\rangle_{AB}=\left| \Theta^- \right\rangle_{AB}$ and $C\left| \Theta^- \right\rangle_{AB}=\left| \Theta^+ \right\rangle_{AB}$. This shows that the separable states $\left| \Theta^\pm \right\rangle_{AB}$ are not the eigenstates of $C$.

Let us now apply the operator $C$ to the superposition states $\left| \Psi^\pm \right\rangle_{AB}$, we have, 
\begin{equation}
\begin{aligned}
\label{SuperpositionStates}	
C\left| \Psi^\pm \right\rangle_{AB} &= \frac{1}{\sqrt{2}} \left[C\left(\left| P \right\rangle_A \left| \bar{P} \right\rangle_B \right) \pm C\left(\left| \bar{P} \right\rangle_A \left| P \right\rangle_B  \right)\right], \\	
&= \frac{1}{\sqrt{2}} \left( \left| \bar{P} \right\rangle_A \left| P \right\rangle_B  \pm \left| P \right\rangle_A \left| \bar{P} \right\rangle_B  \right),  \\
&= \pm \left| \Psi^\pm \right\rangle_{AB}.  	 
\end{aligned}
\end{equation}

Eq.(\ref{SuperpositionStates}) shows that the superposition states $\left| \Psi^\pm \right\rangle_{AB}$ are the eigenstates of the charge conjugation operator $C$. On the other hand, one can easily show that $C=C^\dag$.\cite{Hermitian} This means that $C$ is a Hermitian operator and is therefore an observable physical quantity. As the eigenstates of $C$, therefore, $\left| \Psi^\pm \right\rangle_{AB}$ must exist.

The separable states $\left| \Theta^\pm \right\rangle_{AB}$ have a fundamental character: each of their particles is either a particle or an antiparticle, the particles' identities are determinate. But the superposition states $\left| \Psi^\pm \right\rangle_{AB}$ (eigenstates of $C$) are entangled states because they cannot be expressed as the direct product of the particle state and antiparticle state.\cite{Horodecki,Lo} A fundamental character of $\left| \Psi^\pm \right\rangle_{AB}$ is that each of their particles is a superposition of a particle and an antiparticle, i.e., it is partially a particle and partially an antiparticle. Therefore, one cannot tell which one is a particle and which one is an antiparticle.
When performing a measurement on $A$, it will collapse into either a particle $\left| P \right\rangle_A$, or an antiparticle $\left| \bar{P} \right\rangle_A$. If $A$ collapse into a particle $\left| P \right\rangle_A$, then $B$ will collapse into an antiparticle $\left| \bar{P} \right\rangle_B$, i.e., $\left| \Psi^\pm \right\rangle_{AB} \rightarrow \left| P \right\rangle_A\left| \bar{P} \right\rangle_B $. If $A$ collapse into an antiparticle $\left| \bar{P} \right\rangle_A$, then $B$ will collapse into a particle $\left| P \right\rangle_B$, i.e., $\left| \Psi^\pm \right\rangle_{AB} \rightarrow \pm \left| \bar{P} \right\rangle_A\left| P \right\rangle_B$.

As mentioned before, the charge conjugation operator $C$ packages a number of quantum numbers ($Q$, $B$, $L$, $I_3$, $C$, $S$, $T$, $B'$). All these quantum numbers should be entangled together in the packaged entangled states $\left| \Psi^\pm \right\rangle_{AB}$. In other words, $\left| \Psi^\pm \right\rangle_{AB}$ cannot be written as the product of the sub quantum states related to the individual quantum numbers (freedoms). This feature can be embodied by rewriting Eq.(\ref{PsiE}), i.e.,
\begin{equation}
\begin{aligned}
\label{PackagedEntangle}
\left| \Psi^\pm \right\rangle_{AB} =& \frac{1}{\sqrt{2}} \left( \left| Q, B, L, I_3, C, S, T, B' \right\rangle_A \left| -Q, -B, -L, -I_3, -C, -S, -T, -B' \right\rangle_B \right.  \\
& \left. \pm  \left| -Q, -B, -L, -I_3, -C, -S, -T, -B' \right\rangle_A \left| Q, B, L, I_3, C, S, T, B' \right\rangle_B  \right).
\end{aligned}
\end{equation}

In fact, the packaged entangled states $\left| \Psi^\pm \right\rangle_{AB}$ in Eq.(\ref{PsiE}) only include abstract particle symbols, $\left| P \right\rangle_A$, $\left| \bar{P} \right\rangle_A$, $\left| P \right\rangle_B$, $\left| \bar{P} \right\rangle_B$, but not some particular physical quantity symbols. This means that the quantum numbers ($Q$, $B$, $L$, $I_3$, $C$, $S$, $T$, $B'$) is packaged as an entirety in $\left| \Psi^\pm \right\rangle_{AB}$ by the charge conjugation operator $C$. The quantum numbers cannot be added in or taken out separately.
Furthermore, a particle and an antiparticle are different types of particles. It means that $\left| \Psi^\pm \right\rangle_{AB}$ package different particles as an entirety. Due to these reasons, we call the quantum states $\left| \Psi^\pm \right\rangle_{AB}$ as packaged entangled states.

\subsection{Packaged entangled states with C-symmetry breaking}

The packaged entangled states $\left| \Psi^\pm \right\rangle_{AB}$ in Eq.(\ref{PsiE}) are construct on the basis of a particle-antiparticle pair in which the total charge is conserved (zero). They strictly obey the law of charge conservation in the wave function collapse.
From a mathematical point of view, however, there should be other forms of packaged entangled states in which the total charge are not conserved in the wave function collapse, i.e., the total charge before the wave function collapse is not equal to that after the wave function collapse. Let us now construct the mathematical expressions for these new packaged entangled states.

Consider the following two quantum states of a particle pair,
\begin{subequations}
	\label{PhiE}
	\begin{eqnarray}
	\left| \Phi^+ \right\rangle_{AB} &= \frac{1}{\sqrt{2}} \left(\left| P \right\rangle_A \left| P \right\rangle_B + \left| \bar{P} \right\rangle_A \left| \bar{P} \right\rangle_B  \right), \label{PhiP} \\
	\left| \Phi^- \right\rangle_{AB} &= \frac{1}{\sqrt{2}} \left(\left| P \right\rangle_A \left| P \right\rangle_B - \left| \bar{P} \right\rangle_A \left| \bar{P} \right\rangle_B  \right). \label{PhiM}
	\end{eqnarray}
\end{subequations}

Applying the charge conjugation operator $C$ to $\left| \Phi^\pm \right\rangle_{AB}$, we have
\begin{equation}
\begin{aligned}
\label{OperatorC}
C\left| \Phi^\pm \right\rangle_{AB}
&= \frac{1}{\sqrt{2}} \left[ C \left(\left| P \right\rangle_A \left| P \right\rangle_B \right) \pm C \left(\left| \bar{P} \right\rangle_A \left| \bar{P} \right\rangle_B \right)  \right],  \\
&= \frac{1}{\sqrt{2}} \left[ \left| \bar{P} \right\rangle_A \left| \bar{P} \right\rangle_B \pm \left| P \right\rangle_A \left| P \right\rangle_B \right]  \\
&= \pm \left| \Phi^\pm \right\rangle_{AB}.
\end{aligned}
\end{equation}

Eq.(\ref{OperatorC}) shows that $\left| \Phi^\pm \right\rangle_{AB}$ are also the eigenstates of charge conjugation operator $C$. Because $C$ is a Hermitian operator (or an observable physical quantity) \cite{Hermitian}, its eigenstates $\left| \Phi^\pm \right\rangle_{AB}$ must exist.

Similar to $\left| \Psi^\pm \right\rangle_{AB}$, the states $\left| \Phi^\pm \right\rangle_{AB}$ are also entangled states because they cannot be expressed as the direct product of the particle states and antiparticle states.\cite{Horodecki,Lo} Furthermore, as the eigenstates of the charge conjugation operator $C$, the entangled states $\left| \Phi^\pm \right\rangle_{AB}$ also package in all the physical properties capable of completely identifying the particles, i.e., the particle's electric charge ($Q$), baryon number ($B$), lepton number ($L$), isospin ($I_3$), charm ($C$), strangeness ($S$), topness ($T$), and bottomness ($B'$).

The packaged entangled states $\left| \Phi^\pm \right\rangle_{AB}$ have an interesting property. If a measurement is performed on the particle pair, $\left| \Phi^+ \right\rangle_{AB}$ (or $\left| \Phi^- \right\rangle_{AB}$) will collapse and break the C-symmetry (the symmetry of physical laws under the charge conjugation operator $C$) \cite{Griffiths,Perkins,Peskin}. More specifically, if a measurement is performed on $A$, it will collapse into either a particle $\left| P \right\rangle_A$, or an antiparticle $\left| \bar{P} \right\rangle_A$.
If $A$ collapse into a particle $\left| P \right\rangle_A$, then $B$ will also collapse into a particle $\left| P \right\rangle_B$, i.e., $\left| \Phi^\pm \right\rangle_{AB} \rightarrow \left| P \right\rangle_A \left| P \right\rangle_B$.  
If $A$ collapse into an antiparticle $\left| \bar{P} \right\rangle_A$, then $B$ will also collapse into an antiparticle $\left| \bar{P} \right\rangle_B$, i.e., $\left| \Phi^\pm \right\rangle_{AB} \rightarrow \pm \left| \bar{P} \right\rangle_A \left| \bar{P} \right\rangle_B$. 
This process break the C-symmetry of the particle-antiparticle pair. Therefore, the law of charge conservation does not hold in this process.

\subsection{Annihilation of particles in packaged entangled states}
\label{Annihi}

The exact experimental methods for generating the packaged entangled states are unavailable at present. However, it is well known that a particle and an antiparticle will annihilate each other when they encounter.\cite{Leinaas} Thus, one may ask what will happen when an external particle (from a particle source) encounters a particle in a packaged entangled state?

Recall that external perturbation or a measurement on the particles in a packaged entangled state will cause the wave function to collapse. 
Because each particle in the packaged entangled state is a superposition of a particle and an antiparticle, the collapse of the packaged entangled state is then not random, but has a partiality depending on the external particle due to the particle-antiparticle annihilation phenomenon.
In other words, when an external particle $X$ (from a particle source) encounters a particle $A$ that is in a packaged entangled state, the particle-antiparticle annihilation phenomenon \cite{Anni,Klempt} will force $A$ to collapse into a particle conjugating to $X$ (with every internal quantum number of $A$ opposite to that of $X$), or project $A$ onto a state conjugating to that of $X$. Thereafter, $X$ and $A$ annihilate each other. This is not a process of continuous evolution via the Schr\"{o}dinger equation.\cite{Schlosshauer} More specifically,

1. If $X$ is a particle (denoted as $\left| P \right\rangle_X$), then $A$ will collapse into an antiparticle $\left| \bar{P} \right\rangle_A$. The collapse of total wave function may be expressed as, 
\[\left| P \right\rangle_X \left| \Psi^\pm \right\rangle_{AB} ~ \longrightarrow ~ \pm \left( \left| P \right\rangle_X \left| \bar{P} \right\rangle_A \right) \left| P \right\rangle_B, \]
\[\left| P \right\rangle_X \left| \Phi^\pm \right\rangle_{AB} ~ \longrightarrow ~ \pm \left( \left| P \right\rangle_X \left| \bar{P} \right\rangle_A \right) \left| \bar{P} \right\rangle_B; \]

2. If $X$ is an antiparticle (denoted as $\left| \bar{P} \right\rangle_X$), then $A$ will collapse into a particle $\left| P \right\rangle_A$. The collapse of total wave function may be expressed as,
\[\left| \bar{P} \right\rangle_X \left| \Psi^\pm \right\rangle_{AB} ~ \longrightarrow ~ \left( \left| \bar{P} \right\rangle_X \left| P \right\rangle_A \right) \left| \bar{P} \right\rangle_B, \]
\[\left| \bar{P} \right\rangle_X \left| \Phi^\pm \right\rangle_{AB} ~ \longrightarrow ~ \left( \left| \bar{P} \right\rangle_X \left| P \right\rangle_A \right) \left| P \right\rangle_B. \]

One can see that each particle in the packaged entangled states can annihilate with both a particle and an antiparticle. This property could be used for testing the existence of packaged entangled states, particle-antiparticle teleportation, and transfer of packaged entangled states.

\subsection{Generalization to $M (> 2)$-particle systems}
\label{Generalization}

The packaged entangled states constructed in Eq.(\ref{PsiE}) and Eq.(\ref{PhiE}) are valid for 2-particle systems only. We will now generalize them to $M>2$-particle systems.

The quantum states of the $M (> 2)$-particle system can be classified using the number of antiparticles included in the state. Let us first consider the separable states of this system and denote them by $\left| \Theta \right\rangle_{ij}$, where the suffix $i$ represents the number of antiparticles and $j$ labels the $j$-th combination of $i$ antiparticles in total $M$ particles. The value of $j$ can be $j = 1, \cdots, Q$, where $Q = {M \choose i} = \frac{M(M-1)\cdots(M-i+1)}{i!}$ is the number of $i$-combinations from $M$ elements. Thus, we have
\begin{equation}
\begin{aligned}
\label{MSeparableState}
&\left| \Theta \right\rangle_{i1} = \left| \bar{P} \right\rangle_1 \left| \bar{P} \right\rangle_2 \cdots \left| \bar{P} \right\rangle_{i-1} \left| \bar{P} \right\rangle_{i} \left| P \right\rangle_{i+1}\left| P \right\rangle_{i+2} \cdots \left| P \right\rangle_{M-1} \left| P \right\rangle_M, \\
&\left| \Theta \right\rangle_{i2} = \left| \bar{P} \right\rangle_1 \left| \bar{P} \right\rangle_2 \cdots \left| \bar{P} \right\rangle_{i-1} \left| P \right\rangle_{i} \left| \bar{P} \right\rangle_{i+1}\left| P \right\rangle_{i+2} \cdots \left| P \right\rangle_{M-1} \left| P \right\rangle_M, \\
&\cdots\cdots.
\end{aligned}
\end{equation}

In fact, the separable states in Eq.(\ref{MSeparableState}) can be obtained by doing various permutations to the first one. One can see that these separable states are not symmetrical. We can form the symmetrical state by adding the separable states (all possible permutations), i.e.,
\begin{equation}
\label{SymmetricalState}
\left| \Theta \right\rangle_i = \frac{1}{\sqrt{Q}} \left( \left| \Theta \right\rangle_{i1} + \left| \Theta \right\rangle_{i2} + \left| \Theta \right\rangle_{i3} + \cdots + \left| \Theta \right\rangle_{iQ} \right). 
\end{equation}

Eq.(\ref{SymmetricalState}) shows that $\left| \Theta \right\rangle_i$ is an entangled state, but it is not the eigenstate of charge conjugation operator $C$. Finally, we can obtain the eigenstate of operator $C$ by doing a linear combinations on $\left| \Theta \right\rangle_i$ and its charge conjugation state $\left| \bar{\Theta} \right\rangle_i = C \left| \Theta \right\rangle_i$, i.e.,
\begin{equation}
\begin{aligned}
\label{EigenStateM}
\left| \Phi^\pm \right\rangle_i 
&= \frac{1}{\sqrt{2}} \left( \left| \Theta \right\rangle_i \pm \left| \bar{\Theta} \right\rangle_i \right) \\
&= \frac{1}{\sqrt{2Q}} \left[ \left( \left| \Theta \right\rangle_{i1} + \left| \Theta \right\rangle_{i2} \cdots + \left| \Theta \right\rangle_{iQ} \right) \pm \left( \left| \bar{\Theta} \right\rangle_{i1} + \left| \bar{\Theta} \right\rangle_{i2} + \cdots + \left| \bar{\Theta} \right\rangle_{iQ} \right) \right].
\end{aligned}
\end{equation}

One can show that $C \left| \Phi^\pm \right\rangle_i = \frac{1}{\sqrt{2}} \left( \left| \bar{\Theta} \right\rangle_i \pm \left| \Theta \right\rangle_i \right) = \pm \left| \Phi^\pm \right\rangle_i$. This means that the packaged entangled states $\left| \Phi^\pm \right\rangle_i$ are the eigenstates of charge conjugation operator $C$.

It should mentioned that, due to the particle-antiparticle symmetry, the separable states have following symmetry (using the charge conjugation operator $C$),
\begin{equation}
\begin{aligned}
\label{SeparableConjugation}
\left| \bar{\Theta} \right\rangle_{ij} = C \left| \Theta \right\rangle_{ij} = \left| \Theta \right\rangle_{(M-i)j},
\end{aligned}
\end{equation}
where the formula $ {M \choose i} = {M \choose M-i} $ is used.\cite{Zwillinger}
Using Eq.(\ref{SeparableConjugation}), one can easily write out the separable states of the systems with more than $\left[\frac{M}{2}\right]$ (here $\left[\frac{M}{2}\right]$ denotes the integer part of $\frac{M}{2}$) antiparticles (i.e., $M-i$ antiparticles) as follows:
\begin{equation}
\begin{aligned}
\label{PackagedConjugation2}
&\left| \bar{\Theta} \right\rangle_{01} =  \left| \Theta \right\rangle_{M1}, \\
&\left| \bar{\Theta} \right\rangle_{1j} =  \left| \Theta \right\rangle_{(M-1)j}, \\
&\left| \bar{\Theta} \right\rangle_{2j} =  \left| \Theta \right\rangle_{(M-2)j}, \\
&\cdots\cdots, \\
&\left| \bar{\Theta} \right\rangle_{\left[\frac{M}{2}\right]j} = 
\left\{\begin{array}{ll} 
\left| \Theta \right\rangle_{\left[\frac{M}{2}\right]j}, M ~\text{is an even number}, \\
\left| \Theta \right\rangle_{\left(\left[\frac{M}{2}\right]+1\right)j}, M ~\text{is an odd number.}
\end{array} \right.
\end{aligned}
\end{equation}

Let us now explicitly write out some special cases as follows:

\textbf{i). Zero antiparticle} ($i=0$).

The total number of such separable states is ${M \choose 0} = 1$.
\[ \left| \Theta \right\rangle_{01} = \left| P \right\rangle_1 \left| P \right\rangle_2 \left| P \right\rangle_3 \cdots \left| P \right\rangle_{M-1} \left| P \right\rangle_M. \]

The packaged entangled states are (see Eq.(\ref{EigenStateM})) \cite{GHZ}
\begin{equation}
\label{ZeroAntiPart}
\left| \Phi^\pm \right\rangle_0 = \frac{1}{\sqrt{2}} \left( \left| \Theta \right\rangle_{01} \pm \left| \bar{\Theta} \right\rangle_{01} \right).
\end{equation}

If $M=2$, then Eq.(\ref{ZeroAntiPart}) reduce to Eq.(\ref{PhiE}).

\textbf{ii). One antiparticle} ($i=1$).

The total number of such separable states is ${M \choose 1} = M$.
\[ \left| \Theta \right\rangle_{11} = \left| \bar{P} \right\rangle_1 \left| P \right\rangle_2 \left| P \right\rangle_3 \cdots \left| P \right\rangle_{M-1} \left| P \right\rangle_M, \]
\[ \left| \Theta \right\rangle_{12} = \left| P \right\rangle_1 \left| \bar{P} \right\rangle_2 \left| P \right\rangle_3 \cdots \left| P \right\rangle_{M-1} \left| P \right\rangle_M, \]
\[ \cdots\cdots, \]
\[ \left| \Theta \right\rangle_{1M} = \left| P \right\rangle_1 \left| P \right\rangle_2 \left| P \right\rangle_3 \cdots \left| P \right\rangle_{M-1} \left| \bar{P} \right\rangle_M. \]

The symmetrical state is
\begin{equation}
\label{SymmetricalStateOne}
\left| \Theta \right\rangle_1 = \frac{1}{\sqrt{M}} \left( \left| \Theta \right\rangle_{11} + \left| \Theta \right\rangle_{12} + \left| \Theta \right\rangle_{13} + \cdots + \left| \Theta \right\rangle_{1M} \right). 
\end{equation}
and packaged entangled states are
\begin{equation}
\label{OneAntiPart}
\left| \Phi^\pm \right\rangle_1 = \frac{1}{\sqrt{2}} \left( \left| \Theta \right\rangle_1 \pm \left| \bar{\Theta} \right\rangle_1 \right).
\end{equation}

If $M=2$, then Eq.(\ref{OneAntiPart}) reduce to Eq.(\ref{PsiE}).

\textbf{iii). Two antiparticles} ($i=2$). 

The total number of such separable states is $B = {M \choose 2} = \frac{M(M-1)}{2!}$.
\[ \left| \Theta \right\rangle_{21} = \left| \bar{P} \right\rangle_1 \left| \bar{P} \right\rangle_2 \left| P \right\rangle_3 \cdots \left| P \right\rangle_{M-1} \left| P \right\rangle_M, \]
\[ \left| \Theta \right\rangle_{22} = \left| \bar{P} \right\rangle_1 \left| P \right\rangle_2 \left| \bar{P} \right\rangle_3 \cdots \left| P \right\rangle_{M-1} \left| P \right\rangle_M, \]
\[\cdots\cdots.\]
\[\cdots\cdots.\]

The symmetrical state is
\begin{equation}
\label{SymmetricalStateTwo}
\left| \Theta \right\rangle_2 = \frac{1}{\sqrt{B}} \left( \left| \Theta \right\rangle_{21} + \left| \Theta \right\rangle_{22} + \left| \Theta \right\rangle_{23} + \cdots + \left| \Theta \right\rangle_{2B} \right). 
\end{equation}
and packaged entangled states are
\begin{equation}
\label{TwoAntiPart}
\left| \Phi^\pm \right\rangle_2 = \frac{1}{\sqrt{2}} \left( \left| \Theta \right\rangle_2 \pm \left| \bar{\Theta} \right\rangle_2 \right).
\end{equation}

\textbf{iv). $\left[\frac{M}{2}\right]$ antiparticles} ($i=\left[\frac{M}{2}\right]$).

Let $\left[\frac{M}{2}\right]$ denote the integer part of $\frac{M}{2}$. If $M$ is an even number, then $\left[\frac{M}{2}\right]=\frac{M}{2}$. If $M$ is an odd number, then $C = \frac{M}{2}-1 < \left[\frac{M}{2}\right] < \frac{M}{2}$.
The total number of such separable states is ${M \choose \left[M/2\right]} = \frac{M(M-1)\cdots(M-\left[M/2\right]+1)}{\left[M/2\right]!}$.
\[ \left| \Theta \right\rangle_{\left[\frac{M}{2}\right]1} = \left| \bar{P} \right\rangle_1 \left| \bar{P} \right\rangle_2 \cdots \left| \bar{P} \right\rangle_{\left[\frac{M}{2}\right]-1} \left| \bar{P} \right\rangle_{\left[\frac{M}{2}\right]} \left| P \right\rangle_{\left[\frac{M}{2}\right]+1}\left| P \right\rangle_{\left[\frac{M}{2}\right]+2} \cdots \left| P \right\rangle_{M-1} \left| P \right\rangle_M, \]
\[ \left| \Theta \right\rangle_{\left[\frac{M}{2}\right]2} = \left| \bar{P} \right\rangle_1 \left| \bar{P} \right\rangle_2 \cdots \left| \bar{P} \right\rangle_{\left[\frac{M}{2}\right]-1} \left| P \right\rangle_{\left[\frac{M}{2}\right]} \left| \bar{P} \right\rangle_{\left[\frac{M}{2}\right]+1}\left| P \right\rangle_{\left[\frac{M}{2}\right]+2} \cdots \left| P \right\rangle_{M-1} \left| P \right\rangle_M, \]
\[\cdots\cdots.\]

The symmetrical state is
\begin{equation}
\label{SymmetricalStateN}
\left| \Theta \right\rangle_{\left[\frac{M}{2}\right]} = \frac{1}{\sqrt{C}} \left( \left| \Theta \right\rangle_{\left[\frac{M}{2}\right]1} +\left| \Theta \right\rangle_{\left[\frac{M}{2}\right]2} + \cdots + \left| \Theta \right\rangle_{\left[\frac{M}{2}\right]C} \right). 
\end{equation}
and packaged entangled states are
\begin{equation}
\label{HalfMAntiPart}
\left| \Phi^\pm \right\rangle_{\left[\frac{M}{2}\right]} = \frac{1}{\sqrt{2}} \left( \left| \Theta \right\rangle_{\left[\frac{M}{2}\right]} \pm \left| \bar{\Theta} \right\rangle_{\left[\frac{M}{2}\right]} \right).
\end{equation}

From above discussions, we see that the packaged entangled states $\left| \Phi^\pm \right\rangle_i$ have the following properties (Let $N_P$ be the number of $P$s and $N_{\bar{P}}$ be the number of $\bar{P}$s in $\left| \Theta \right\rangle_i$):

1. The collapse of the packaged entangled states $\left| \Phi^\pm \right\rangle_i$ will reduce them to one of the separable states $\left| \Theta \right\rangle_{ij}$ (see Eq.(\ref{MSeparableState})) or $\pm \left| \bar{\Theta} \right\rangle_{ij}$, but not the entangled state $\left| \Theta \right\rangle_i$ (see Eq.(\ref{SymmetricalState})) or $\pm \left| \bar{\Theta} \right\rangle_i$.

2. If $M$ (the number of total particles in the system) is an even number and $N_P=N_{\bar{P}}$, then the C-symmetry holds in the collapse of these wave functions. 
For example, if a measurement is performed on $\left| \Phi^\pm \right\rangle_{\left[\frac{M}{2}\right]}$ (see Eq.(\ref{HalfMAntiPart})), then they will either collapse into one of the separable states $\left| \Theta \right\rangle_{\left[\frac{M}{2}\right]j}$ or $\pm \left| \bar{\Theta} \right\rangle_{\left[\frac{M}{2}\right]j}$. The C-symmetry holds in these processes.

3. If $M$ is an even number but $N_P \neq N_{\bar{P}}$, then the C-symmetry does not hold in the collapse of $\left| \Phi^\pm \right\rangle_i$. For example, if a measurement is performed on $\left| \Phi^\pm \right\rangle_0$ (see Eq.(\ref{ZeroAntiPart})), then they will either collapse into the separable states $\left| \Theta \right\rangle_{01}$ or $\pm \left| \bar{\Theta} \right\rangle_{01}$. These processes break the C-symmetry. Therefore, the C-symmetry does not hold in the collapse of these wave functions.

4. If $M$ is an odd number, then we always have $N_P \neq N_{\bar{P}}$. Therefore, all the packaged entangled states have C-symmetry breaking.

\section{Applications}

\subsection{Particle-antiparticle teleportation with packaged entangled states}

We shall now discuss a possible particle-antiparticle teleportation protocol using the packaged entangled states (see Fig.1). 
Here the particle-antiparticle teleportation does not mean to transmit a particle to a receiver (from Alice to Bob), but only transmit the packaged quantum information carried by the particle to the receiver.\cite{Krauss} 
In the teleportation process, we need to use particle-antiparticle annihilation phenomenon which can be efficiently described by second quantization formalism. But this is not the main point of the teleportation protocol. Our main point is about quantum entanglement and wave function collapse which are related to first quantization.

Let us choose the packaged entangled state $\left| \Psi^+ \right\rangle_{AB}$ in Eq.(\ref{PsiP}) to carry out the calculation. One will see that the sender can control the receiver's particle to be a particle (or an antiparticle) in the teleportation process. For the convenience of discussion, the particle-antiparticle teleportation protocol is divided into 5 steps.

\begin{figure}[htb]
	\label{figure1}
	\begin{center}
		\includegraphics[height=0.45\textwidth]{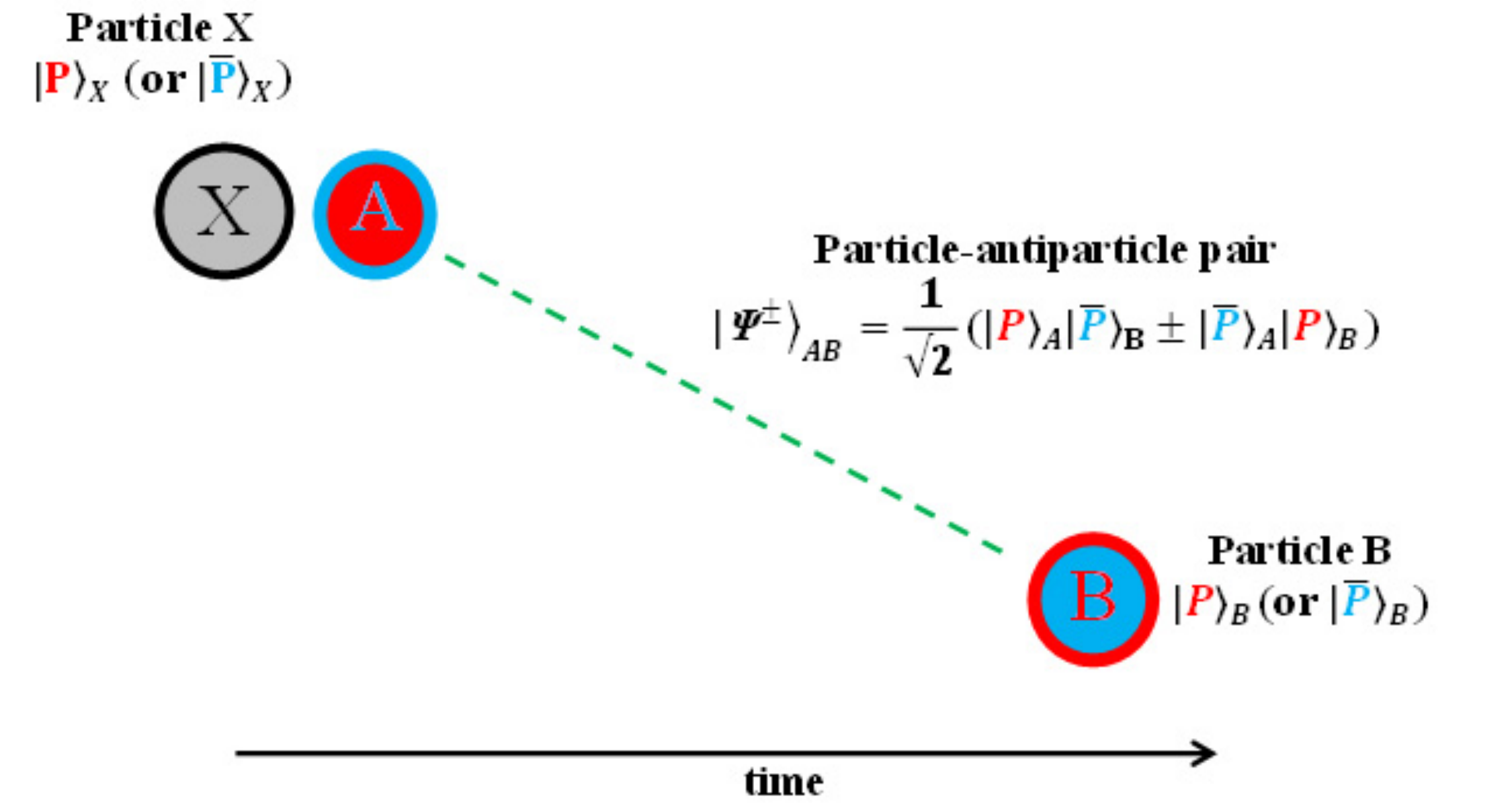}
		\caption{(Color online) Schematic diagram for particle-antiparticle teleportation using the packaged entangled states of a particle-antiparticle pair, $\left| \Psi^\pm \right\rangle_{AB} = \frac{1}{\sqrt{2}} \left( \left| P \right\rangle_A \left| \bar{P} \right\rangle_B \pm  \left| \bar{P} \right\rangle_A \left| P \right\rangle_B  \right)$, and particle-antiparticle annihilation phenomenon.
		}
	\end{center}
\end{figure}

\textbf{(1) Encoding}. 
To carry out a particle-antiparticle teleportation, Alice needs to encode her information in the form of quantum states of particles and antiparticles. Consider that Alice has a particle $X$ (or a sequence of particles) want to teleport to Bob. $X$ is either a particle $\left| P \right\rangle_X$ or an antiparticle $\left| \bar{P} \right\rangle_X$. Without losing generality, let us write out $X$'s quantum state in a single formula as
\begin{equation}
\label{Messenger}
\left| \phi \right\rangle_X = \alpha \left| P \right\rangle_X + \beta \left| \bar{P} \right\rangle_X,
\end{equation}
by assuming that, if $X$ is a particle, then $\alpha=1$ and $\beta=0$; if $X$ is an antiparticle, then $\alpha=0$ and $\beta=1$.

It should be emphasized that currently no physical theories or experiments has shown that a single particle can be in the general superposition state with $\alpha^2+\beta^2=1$ in Eq.(\ref{Messenger}). It is either a particle or an antiparticle (say an electron or a positron). Thus, the quantum information should be a particle-antiparticle sequence as $\dots, \left| P \right\rangle, \dots, \left| \bar{P} \right\rangle, \dots$.

\textbf{(2) Quantum channel creation}.
Alice needs a quantum channel, a particle-antiparticle pair in the packaged entangled states, to send the information stored on particle $X$ to Bob. According to the agreement between Alice and Bob, they choose $\left| \Psi^+ \right\rangle_{AB} = \frac{1}{\sqrt{2}} \left(\left| P \right\rangle_A \left| \bar{P} \right\rangle_B + \left| \bar{P} \right\rangle_A \left| P \right\rangle_B  \right)$ (see Eq.(\ref{PsiP})) to carry out the teleportation.

After the particle-antiparticle pair is created in $\left| \Psi^+ \right\rangle_{AB}$, one of them (particle $A$) is sent to Alice and the other one (particle $B$) is sent to Bob. Before Alice carry out any further operation, the complete state of the three particles ($X$, $A$, $B$) is 
\begin{equation}
\begin{aligned}
\label{XAB}
&\left| \phi \right\rangle_X \left| \Psi^+ \right\rangle_{AB} \\
&=\frac{\alpha}{\sqrt{2}} \left( \left|P\right\rangle_X \left|P\right\rangle_A \left|\bar{P}\right\rangle_B + \left|P\right\rangle_X \left|\bar{P}\right\rangle_A \left|P\right\rangle_B  \right) 
+ \frac{\beta}{\sqrt{2}} \left( \left|\bar{P}\right\rangle_X \left|P\right\rangle_A \left|\bar{P}\right\rangle_B + \left|\bar{P}\right\rangle_X \left|\bar{P}\right\rangle_A \left|P\right\rangle_B  \right).
\end{aligned}
\end{equation}

\textbf{(3) Sending}. 
Alice can send out her information stored on particle $X$ by annihilating particle $X$ with particle $A$. Referring to Section \ref{Annihi} we know that, when particle $X$ encounters particle $A$, the particle-antiparticle annihilation phenomenon \cite{Anni,Klempt} will force $A$ to collapse into a particle conjugating to $X$, or project $A$ onto the state $\left| \bar{\phi} \right\rangle_A = \bar{\alpha} \left| \bar{P} \right\rangle_A + \bar{\beta} \left| P \right\rangle_A$ (see Eq.(\ref{Messenger})). Thereafter, $X$ and $A$ annihilate each other. 
Meanwhile, the three particle state $\left| \phi \right\rangle_X \left| \Psi^+ \right\rangle_{AB}$ (see Eq.(\ref{XAB})) should collapse into a final state $\left| \Psi^+ \right\rangle_{XAB}'$ that only has terms including $\left|P\right\rangle_X \left|\bar{P}\right\rangle_A$ and $\left|\bar{P}\right\rangle_X \left|P\right\rangle_A$ (particle-antiparticle annihilation). One can see that only the second term and third term in Eq.(\ref{XAB}) satisfy these conditions. Therefore, we obtain the final state,
\begin{equation}
\begin{aligned}
\label{XAB2}
\left| \Psi^+ \right\rangle_{XAB}' 
&= \alpha \left( \left|P\right\rangle_X \left|\bar{P}\right\rangle_A \right) \left|P\right\rangle_B + \beta \left( \left|\bar{P}\right\rangle_X \left|P\right\rangle_A \right) \left|\bar{P}\right\rangle_B \\
&= \alpha \left|P\bar{P}\right\rangle_{XA} \left|P\right\rangle_B + \beta \left|\bar{P}P\right\rangle_{XA} \left|\bar{P}\right\rangle_B.
\end{aligned}
\end{equation}
where $\left|P\bar{P}\right\rangle_{XA}$ and $\left|\bar{P}P\right\rangle_{XA}$ are the particles produced by the $\left|P\right\rangle_X \left|\bar{P}\right\rangle_A$ and $\left|\bar{P}\right\rangle_X \left|P\right\rangle_A$ annihilation \cite{Anni}, respectively.

The reduction $\left| \phi \right\rangle_X \left| \Psi^+ \right\rangle_{AB} ~ \longrightarrow ~ \left| \Psi^+ \right\rangle_{XAB}'$ indicates that the state of particle $B$ will be modulated after Alice annihilated particle $X$ and $A$. Thus, Alice successfully sent out her information (state of particle $X$) to Bob (state of particle $B$).

\textbf{(4) Receiving}. 
Bob can receive the packaged information by measuring the state of particle $B$. Recall that before Alice annihilate particle $X$ with $A$, Bob's particle $B$ was in the packaged entangled state $\left| \Psi^+ \right\rangle_{AB}$ and was unrelated to particle $X$. After Alice annihilated particle $X$ and $A$, however, particle $B$ collapsed into a separable state identical to that of particle $X$. This can be seen by by referring to Eq.(\ref{XAB2}). Thus, the teleportation process is a simple correspondence between $X$ and $B$, i.e.,
\[ \left( \alpha \left| P \right\rangle_X + \beta \left| \bar{P} \right\rangle_X \right) ~\sim~ \alpha \left|P\bar{P}\right\rangle_{XA} \left|P\right\rangle_B + \beta \left|\bar{P}P\right\rangle_{XA} \left|\bar{P}\right\rangle_B. \]

This shows that Bob's particle $B$ becomes identical to $X$ after Alice sent out her information by annihilating $X$ with $A$. Therefore, Bob can receive the packaged information of particle $X$ by measuring the state of particle $B$.

\textbf{(5) Decoding}.
Referring to Eq.(\ref{Messenger}) and Eq.(\ref{XAB2}), Bob can decode the packaged information  sent to him by Alice (carried by particle $X$). More specifically, if $X$ is a particle, i.e., $\left|\phi\right\rangle_X = \left|P\right\rangle_X$ (see Eq.(\ref{Messenger})), then Eq.(\ref{XAB2}) becomes $\left| \Psi^+ \right\rangle_{XAB}' = \left|P\bar{P}\right\rangle_{XA} \left|P\right\rangle_B$ and $B$ becomes a particle identical to $X$; If $X$ is an antiparticle, i.e., $\left| \phi \right\rangle_X = \left| \bar{P} \right\rangle_X$, then Eq.(\ref{XAB2}) becomes
$ \left| \Psi^+ \right\rangle_{XAB}' = \left|\bar{P}P\right\rangle_{XA} \left|\bar{P}\right\rangle_B$ and $B$ becomes an antiparticle identical to $X$.
The exact correspondence between $X$ and $B$ can be written as: $\left| P \right\rangle_X ~\sim~ \left|P\right\rangle_B$, $\left| \bar{P} \right\rangle_X ~\sim~ \left|\bar{P}\right\rangle_B$.

Thus, if Bob measure $B$ and find it is a particle, then he can infer that $X$ is a particle. However, if Bob measure $B$ and find it is an antiparticle, then he can infer that $X$ is an antiparticle. In this sense, Bob can successfully decode the packaged information sent to him by Alice (carried by $X$).

Similarly, one can repeat the above particle teleportation process using the other packaged entangled states:

1. If one choose $\left|\Psi^-\right\rangle_{AB}$ (see Eq.(\ref{PsiM})), then Eq.(\ref{XAB2}) becomes
\begin{equation}
\begin{aligned}
\label{XAB5}
\left| \Psi^- \right\rangle_{XAB}' = - \alpha \left|P\bar{P}\right\rangle_{XA} \left|P\right\rangle_B  + \beta \left|\bar{P}P\right\rangle_{XA} \left|\bar{P}\right\rangle_B.
\end{aligned}
\end{equation}

If $\left|\phi\right\rangle_X = \left|P\right\rangle_X$, then $\left| \Psi^- \right\rangle_{XAB}' = - \left|P\bar{P}\right\rangle_{XA} \left|P\right\rangle_B$. If $\left| \phi \right\rangle_X = \left| \bar{P} \right\rangle_X$, then $\left| \Psi^- \right\rangle_{XAB}' = \left|\bar{P}P\right\rangle_{XA} \left|\bar{P}\right\rangle_B$.

2. If one choose $\left| \Phi^+ \right\rangle_{AB}$ (see Eq.(\ref{PhiE})), then Eq.(\ref{XAB2}) becomes
\begin{equation}
\begin{aligned}
\label{XAB2B}
\left| \Phi^+ \right\rangle_{XAB}' 
&= \alpha \left|P\bar{P}\right\rangle_{XA} \left|\bar{P}\right\rangle_B  + \beta \left|\bar{P}P\right\rangle_{XA} \left|P\right\rangle_B.
\end{aligned}
\end{equation}

If $\left|\phi\right\rangle_X = \left|P\right\rangle_X$, then $\left| \Phi^+ \right\rangle_{XAB}' = \left|P\bar{P}\right\rangle_{XA} \left|\bar{P}\right\rangle_B$ ($B$ becomes an antiparticle conjugating to $X$). 
If $\left| \phi \right\rangle_X = \left| \bar{P} \right\rangle_X$, then $\left| \Phi^+ \right\rangle_{XAB}' = \left|\bar{P}P\right\rangle_{XA} \left|P\right\rangle_B$ ($B$ becomes an particle conjugating to $X$).

3. If one choose $\left|\Phi^-\right\rangle_{AB}$ (see Eq.(\ref{PhiE})), then Eq.(\ref{XAB2}) becomes
\begin{equation}
\begin{aligned}
\label{XAB5B}
\left| \Phi^- \right\rangle_{XAB}' = - \alpha \left|P\bar{P}\right\rangle_{XA} \left|\bar{P}\right\rangle_B  + \beta \left|\bar{P}P\right\rangle_{XA} \left|P\right\rangle_B.
\end{aligned}
\end{equation}

If $\left|\phi\right\rangle_X = \left|P\right\rangle_X$, then $\left| \Phi^- \right\rangle_{XAB}' = - \left|P\bar{P}\right\rangle_{XA} \left|\bar{P}\right\rangle_B$. If $\left| \phi \right\rangle_X = \left| \bar{P} \right\rangle_X$, then $\left| \Phi^- \right\rangle_{XAB}' = \left|\bar{P}P\right\rangle_{XA} \left|P\right\rangle_B$.

The above discussion shows that, if Alice and Bob choose $\left| \Psi^\pm \right\rangle_{AB}$ (packaged entangled states with C-symmetry, see Eq.(\ref{PsiE}), then Bob can obtain a particle (particle $B$) identical to that of Alice (particle $X$). However, if Alice and Bob choose $\left| \Phi^\pm \right\rangle_{AB}$ (packaged entangled states with C-symmetry breaking, see Eq.(\ref{PhiE})), then Bob can obtain a particle conjugating to that of Alice.

Using the states $\left| \Psi^\pm \right\rangle_{AB}$, the particle-antiparticle teleportation process satisfies a number of conservation principles, i.e., charge (Q) conservation, baryon number ($B$) conservation, lepton number ($L$) conservation, isospin ($I_3$) conservation, charm ($C$) conservation, strangeness ($S$) conservation, topness ($T$) conservation, and bottomness ($B'$) conservation. 
Using the states $\left| \Phi^\pm \right\rangle_{AB}$ , however, the particle-antiparticle teleportation process does not satisfy the above conservation principles.
But in both cases, the particle-antiparticle teleportation processes satisfy the principles of linear momentum conservation, total energy conservation, and angular momentum conservation.

Finally, let us discuss how to teleport particles to multiple receivers \cite{Dur,Zhao} using one of the packaged entangled states in Eq.(\ref{OneAntiPart}), i.e.,
\begin{equation}
\begin{aligned}
\label{PackagedM2C}
\left| \Phi^+ \right\rangle_1 = \frac{1}{\sqrt{2M}} \left[ \left( \left| \Theta \right\rangle_{11} + \left| \Theta \right\rangle_{12} + \cdots + \left| \Theta \right\rangle_{1M} \right) + \left( \left| \bar{\Theta} \right\rangle_{11} + \left| \bar{\Theta} \right\rangle_{12} + \cdots + \left| \bar{\Theta} \right\rangle_{1M} \right) \right].
\end{aligned}
\end{equation}

Consider that Alice has a particle $X$ as described by Eq.(\ref{Messenger}) and wants to teleport it to the $M-1$ receivers: Bob, Carl, David, Edward, Frank, $\cdots$. First, send particle $1$ in the packaged entangled states $\left| \Phi^+ \right\rangle_1$ to Alice and send the other $M-1$ particles to the multiple receivers, respectively. Before Alice carry out any further operation, the complete state of the $1+M$ particles (see Eq.(\ref{XAB})) is 
\begin{equation}
\begin{aligned}
\label{XAM}
\left| \phi \right\rangle_X \left| \Phi^+ \right\rangle_1 
&=\frac{\alpha}{\sqrt{2M}} \left| P \right\rangle_X \left[ \left( \left| \Theta \right\rangle_{11} + \left| \Theta \right\rangle_{12} + \cdots + \left| \Theta \right\rangle_{1M} \right) + \left( \left| \bar{\Theta} \right\rangle_{11} + \left| \bar{\Theta} \right\rangle_{12} + \cdots + \left| \bar{\Theta} \right\rangle_{1M} \right) \right] \\
&+ \frac{\beta}{\sqrt{2M}} \left| \bar{P} \right\rangle_X \left[ \left( \left| \Theta \right\rangle_{11} + \left| \Theta \right\rangle_{12} + \cdots + \left| \Theta \right\rangle_{1M} \right) + \left( \left| \bar{\Theta} \right\rangle_{11} + \left| \bar{\Theta} \right\rangle_{12} + \cdots + \left| \bar{\Theta} \right\rangle_{1M} \right) \right].
\end{aligned}
\end{equation}

Thereafter, Alice sends out her information stored on particle $X$ by annihilating particle $X$ with particle $1$. The $1+M$ particle state $\left| \phi \right\rangle_X \left| \Phi^+ \right\rangle_1$ (see Eq.(\ref{XAM})) should collapse into a final state $\left| \Phi^+ \right\rangle_{X1(M-1)}'$ that only has terms including $\left|P\right\rangle_X \left|\bar{P}\right\rangle_1$ and $\left|\bar{P}\right\rangle_X \left|P\right\rangle_1$ (particle-antiparticle annihilation). One can see that only the terms $\left| \Theta \right\rangle_{11}$ and $\left| \bar{\Theta} \right\rangle_{11}$ in Eq.(\ref{XAM}) satisfy these conditions. Therefore, we obtain the final state,
\begin{equation}
\begin{aligned}
\label{XAM2}
\left| \Phi^+ \right\rangle_{X1(M-1)}'
&=\alpha \left|P\right\rangle_X \left| \Theta \right\rangle_{11} + \beta \left|\bar{P}\right\rangle_X \left| \bar{\Theta} \right\rangle_{11}  \\
&= \alpha \left|P\right\rangle_X \left| \bar{P} \right\rangle_1 \left| P \right\rangle_2 \left| P \right\rangle_3 \cdots \left| P \right\rangle_{M-1} \left| P \right\rangle_M + \beta \left|\bar{P}\right\rangle_X \left| P \right\rangle_1 \left| \bar{P} \right\rangle_2 \left| \bar{P} \right\rangle_3 \cdots \left| \bar{P} \right\rangle_{M-1} \left| \bar{P} \right\rangle_M  \\
&=\alpha \left|P\bar{P}\right\rangle_{X1} \left| P \right\rangle_2 \left| P \right\rangle_3 \cdots \left| P \right\rangle_{M-1} \left| P \right\rangle_M  
+ \beta \left|\bar{P}P\right\rangle_{X1} \left| \bar{P} \right\rangle_2 \left| \bar{P} \right\rangle_3 \cdots \left| \bar{P} \right\rangle_{M-1} \left| \bar{P} \right\rangle_M.
\end{aligned}
\end{equation}

If $\left|\phi\right\rangle_X = \left|P\right\rangle_X$, then $\left| \Phi^+ \right\rangle_{X1(M-1)}' = \left|P\bar{P}\right\rangle_{X1} \left| P \right\rangle_2 \left| P \right\rangle_3 \cdots \left| P \right\rangle_{M-1} \left| P \right\rangle_M$. If $\left| \phi \right\rangle_X = \left| \bar{P} \right\rangle_X$, then $\left| \Phi^+ \right\rangle_{X1(M-1)}' = \left|\bar{P}P\right\rangle_{X1} \left| \bar{P} \right\rangle_2 \left| \bar{P} \right\rangle_3 \cdots \left| \bar{P} \right\rangle_{M-1} \left| \bar{P} \right\rangle_M$.

Eq.(\ref{XAM2}) shows that, using $\left| \Phi^+ \right\rangle_1$, each receiver can receive a particle identical to particle $X$. This confirms that Alice can teleport particles to multiple receivers.

\subsection{Transfer of packaged entangled states}

As mentioned before, Alice needs a quantum channel (a particle pair in a packaged entangled state) to perform particle-antiparticle teleportation. However, if two un-entangled particles are spatially separated by large distance, then it is difficult to put these two particles into a packaged entangled state. In this case, one should consider the possibility of transferring the packaged entangled state from the entangled particles to the objective particles which are originally unrelated.

The purpose of this section is to study the entanglement transfer. The procedure is similar but not equal to the entanglement swapping \cite{Zukowski,Schmid}. The fundamental difference is that the entanglement swapping process use Bell measurements to swap the entanglements, but here we will use particle-antiparticle annihilation phenomenon to transfer the packaged entanglements. First, let us choose $\left| \Psi^+ \right\rangle_{AB} = \frac{1}{\sqrt{2}} \left(\left| P \right\rangle_A \left| \bar{P} \right\rangle_B - \left| \bar{P} \right\rangle_A \left| P \right\rangle_B  \right)$ (see Eq.(\ref{PsiP})) to carry out the calculation (see Fig. 2).

Consider that particle $A$ and $B$ are originally in the packaged entangled state $\left| \Psi^+ \right\rangle_{AB}$, and particle $C$ and $D$ are in the packaged entangled state $\left| \Psi^+ \right\rangle_{CD}$, i.e.,
\begin{equation}
\begin{aligned}
\label{PsiABPsiCD}
\left| \Psi^+ \right\rangle_{AB} &=& \frac{1}{\sqrt{2}} \left(\left| P \right\rangle_A \left| \bar{P} \right\rangle_B + \left| \bar{P} \right\rangle_A \left| P \right\rangle_B  \right),  \\
\left| \Psi^+ \right\rangle_{CD} &=& \frac{1}{\sqrt{2}} \left(\left| P \right\rangle_C \left| \bar{P} \right\rangle_D + \left| \bar{P} \right\rangle_C \left| P \right\rangle_D  \right).
\end{aligned}
\end{equation}

\begin{figure}[htb]
	\label{figure2}
	\begin{center}
		\includegraphics[height=0.45\textwidth]{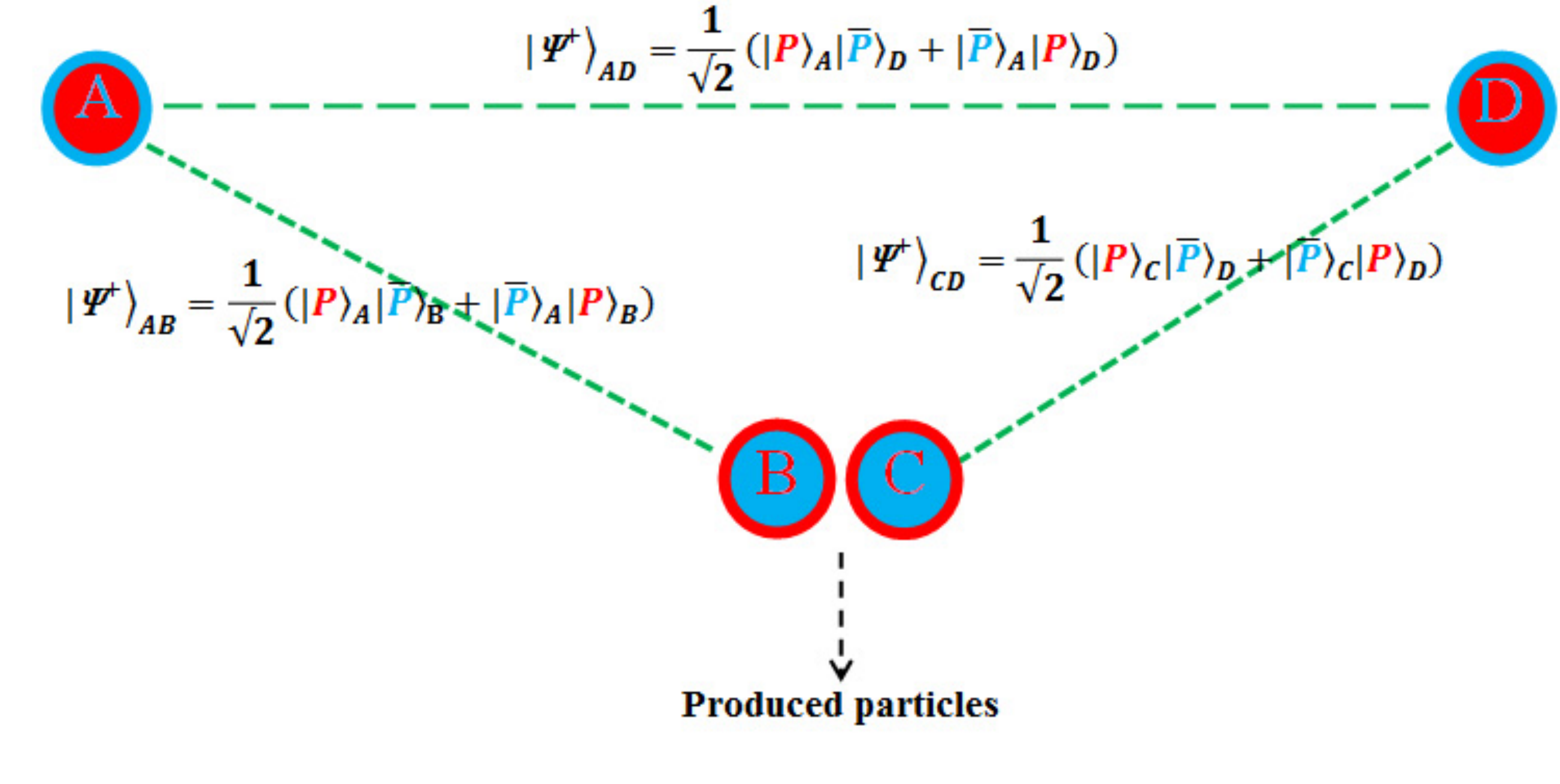}
		\caption{(Color online) Schematic diagram for the entanglement transfer from the packaged entangled states
			$\left| \Psi^+ \right\rangle_{AB} = \frac{1}{\sqrt{2}} \left(\left| P \right\rangle_A \left| \bar{P} \right\rangle_B + \left| \bar{P} \right\rangle_A \left| P \right\rangle_B \right)$ and 
			$\left| \Psi^+ \right\rangle_{CD} = \frac{1}{\sqrt{2}} \left(\left| P \right\rangle_C \left| \bar{P} \right\rangle_D + \left| \bar{P} \right\rangle_C \left| P \right\rangle_D \right)$ to the packaged entangled state
			$\left| \Psi^+ \right\rangle_{AD} = \frac{1}{\sqrt{2}} \left(\left| P \right\rangle_A \left| \bar{P} \right\rangle_D + \left| \bar{P} \right\rangle_A \left| P \right\rangle_D \right)$.
		}
	\end{center}
\end{figure}

Apparently, $A$ and $D$ are unrelated, $B$ and $C$ are unrelated. Now we wish to connect $A$ and $D$ in a packaged entangled state without touching them. This can be achieved by annihilating $B$ and $C$.

Before Alice carry out any further operation, the complete state of the four particles ($A$, $B$, $C$, $D$) is 
\begin{equation}
\begin{aligned}
\label{PsiABCD2}
\left| \Psi^+ \right\rangle_{ABCD} =& \left| \Psi^+ \right\rangle_{AB} \left| \Psi^+ \right\rangle_{CD} \\
=& \frac{1}{2} \left(
\left| P \right\rangle_A \left| \bar{P} \right\rangle_B \left| P \right\rangle_C \left| \bar{P} \right\rangle_D + \left| P \right\rangle_A \left| \bar{P} \right\rangle_B \left| \bar{P} \right\rangle_C \left| P \right\rangle_D \right. \\
&\left. + \left| \bar{P} \right\rangle_A \left| P \right\rangle_B \left| P \right\rangle_C \left| \bar{P} \right\rangle_D + \left| \bar{P} \right\rangle_A \left| P \right\rangle_B \left| \bar{P} \right\rangle_C \left| P \right\rangle_D 
\right) 
\end{aligned}
\end{equation}

Each particle in the packaged entangled states is a superposition of a particle and an antiparticle. When particle $B$ encounters $C$, the particle-antiparticle annihilation phenomenon \cite{Anni,Klempt} will force $B$ and $C$ to collapse into a pair of conjugated particles ($\left|P\right\rangle_B \left|\bar{P}\right\rangle_C$ or $\left|\bar{P}\right\rangle_B \left|P\right\rangle_C$). Afterwards, the particle-antiparticle pair ($B$, $C$) will annihilate each other. Thus, the $\left| \Psi^+ \right\rangle_{ABCD}$ in Eq.(\ref{PsiABCD2}) will collapse into a state $\left| \Psi^{++} \right\rangle_{ABCD}'$ which only has terms including $\left|P\right\rangle_B \left|\bar{P}\right\rangle_C$ and $\left|\bar{P}\right\rangle_B \left|P\right\rangle_C$, i.e.,
\begin{equation}
\begin{aligned}
\label{PsiABCD3}
\left| \Psi^{++} \right\rangle_{ABCD}' 
&= \frac{1}{\sqrt{2}} \left(
\left| P \right\rangle_A \left| \bar{P} \right\rangle_B \left| P \right\rangle_C \left| \bar{P} \right\rangle_D + \left| \bar{P} \right\rangle_A \left| P \right\rangle_B \left| \bar{P} \right\rangle_C \left| P \right\rangle_D 
\right) \\
&=  \left| \Psi^+ \right\rangle_{AD} \left| P \bar{P} \right\rangle_{BC}.
\end{aligned}
\end{equation}
where $\left| \Psi^+ \right\rangle_{AD} = \frac{1}{\sqrt{2}} \left(\left| P \right\rangle_A \left| \bar{P} \right\rangle_D + \left| \bar{P} \right\rangle_A \left| P \right\rangle_D \right)$ and $\left|P\bar{P}\right\rangle_{BC}$ are the particles produced by the $\left|P\right\rangle_B \left|\bar{P}\right\rangle_C$ and $\left|\bar{P}\right\rangle_B \left|P\right\rangle_C$ annihilation \cite{Anni}.

Eq.(\ref{PsiABCD3}) shows that after the annihilation of particle $B$ and $C$, particle $A$ and $D$ (originally unrelated) is now in the packaged entangled states $\left| \Psi^+ \right\rangle_{AD}$.

Furthermore, the above transfer process can be performed in a sequence or chain with any number of packaged entanglement pairs, i.e.,
\[ A - \overbrace{B \cdots C} - \overbrace{D \cdots E} - \overbrace{F \cdots G} - \overbrace{H \cdots I} - J \cdots. \]

Similarly, one can repeat the transfer process using other combinations of packaged entangled states. The results are summarized in Table \ref{tab:TansferedTab}:
\begin{table}[H]
	\caption{Transfer of packaged entangled states}
	\label{tab:TansferedTab}
\begin{tabular}{c|c|c}
\hline \hline
Combinations in	Eq.(\ref{PsiABPsiCD}) &Eq.(\ref{PsiABCD3}) &Transferred state $\left| \dots \right\rangle_{AD}$ \\
\hline
$\left| \Psi^+ \right\rangle_{AB} \left| \Psi^+ \right\rangle_{CD}$ &$\left| \Psi^{++} \right\rangle_{ABCD}'=\left| \Psi^+ \right\rangle_{AD} \left| P \bar{P} \right\rangle_{BC}$ &$\left| \Psi^+ \right\rangle_{AD} = \frac{1}{\sqrt{2}} \left(\left| P \right\rangle_A \left| \bar{P} \right\rangle_D + \left| \bar{P} \right\rangle_A \left| P \right\rangle_D \right)$  \\
\hline
$\left| \Psi^- \right\rangle_{AB}\left| \Psi^- \right\rangle_{CD}$ &$\left| \Psi^{--} \right\rangle_{ABCD}' = \left| \Psi^+ \right\rangle_{AD} \left| P \bar{P} \right\rangle_{BC}$ &$\left| \Psi^+ \right\rangle_{AD} = \frac{1}{\sqrt{2}} \left(\left| P \right\rangle_A \left| \bar{P} \right\rangle_D + \left| \bar{P} \right\rangle_A \left| P \right\rangle_D \right)$  \\
\hline
$\left| \Psi^+ \right\rangle_{AB}\left| \Psi^- \right\rangle_{CD}$ &$\left| \Psi^{+-} \right\rangle_{ABCD}' =  \left| \Psi^- \right\rangle_{AD} \left| P \bar{P} \right\rangle_{BC}$ &$\left| \Psi^- \right\rangle_{AD} = \frac{1}{\sqrt{2}} \left(\left| P \right\rangle_A \left| \bar{P} \right\rangle_D - \left| \bar{P} \right\rangle_A \left| P \right\rangle_D \right)$  \\
\hline
$\left| \Phi^+ \right\rangle_{AB}\left| \Phi^+ \right\rangle_{CD}$ &$\left| \Phi^{++} \right\rangle_{ABCD}' = \left| \Psi^+ \right\rangle_{AD} \left| P \bar{P} \right\rangle_{BC}$ &$\left| \Psi^+ \right\rangle_{AD} = \frac{1}{\sqrt{2}} \left(\left| P \right\rangle_A \left| \bar{P} \right\rangle_D + \left| \bar{P} \right\rangle_A \left| P \right\rangle_D \right)$  \\
\hline
$\left| \Phi^- \right\rangle_{AB}\left| \Phi^- \right\rangle_{CD}$ &$\left| \Phi^{--} \right\rangle_{ABCD}' = -\left| \Psi^+ \right\rangle_{AD} \left| P \bar{P} \right\rangle_{BC}$ &$\left| \Psi^+ \right\rangle_{AD} = \frac{1}{\sqrt{2}} \left(\left| P \right\rangle_A \left| \bar{P} \right\rangle_D + \left| \bar{P} \right\rangle_A \left| P \right\rangle_D \right)$  \\
\hline
$\left| \Phi^+ \right\rangle_{AB}\left| \Phi^- \right\rangle_{CD}$ &$\left| \Phi^{+-} \right\rangle_{ABCD}' = - \left| \Psi^- \right\rangle_{AD} \left| P \bar{P} \right\rangle_{BC}$ &$\left| \Psi^- \right\rangle_{AD} = \frac{1}{\sqrt{2}} \left(\left| P \right\rangle_A \left| \bar{P} \right\rangle_D - \left| \bar{P} \right\rangle_A \left| P \right\rangle_D \right)$  \\
\hline
$\left| \Psi^+ \right\rangle_{AB}\left| \Phi^+ \right\rangle_{CD}$ &$\left| \Omega^{++} \right\rangle_{ABCD}' = \left| \Phi^+ \right\rangle_{AD} \left| P \bar{P} \right\rangle_{BC}$ &$\left| \Phi^+ \right\rangle_{AD} = \frac{1}{\sqrt{2}} \left(\left| P \right\rangle_A \left| P \right\rangle_D + \left| \bar{P} \right\rangle_A \left| \bar{P} \right\rangle_D \right)$  \\
\hline
$\left| \Psi^+ \right\rangle_{AB}\left| \Phi^- \right\rangle_{CD}$ &$\left| \Omega^{+-} \right\rangle_{ABCD}' = \left| \Phi^- \right\rangle_{AD} \left| P \bar{P} \right\rangle_{BC}$ &$\left| \Phi^- \right\rangle_{AD} = \frac{1}{\sqrt{2}} \left(\left| P \right\rangle_A \left| P \right\rangle_D - \left| \bar{P} \right\rangle_A \left| \bar{P} \right\rangle_D \right)$  \\
\hline
$\left| \Psi^- \right\rangle_{AB}\left| \Phi^+ \right\rangle_{CD}$ &$\left| \Omega^{-+} \right\rangle_{ABCD}' = \left| \Phi^- \right\rangle_{AD} \left| P \bar{P} \right\rangle_{BC}$ &$\left| \Phi^- \right\rangle_{AD} = \frac{1}{\sqrt{2}} \left(\left| P \right\rangle_A \left| P \right\rangle_D - \left| \bar{P} \right\rangle_A \left| \bar{P} \right\rangle_D \right)$  \\
\hline
$\left| \Psi^- \right\rangle_{AB}\left| \Phi^- \right\rangle_{CD}$ &$\left| \Omega^{--} \right\rangle_{ABCD}' = \left| \Phi^+ \right\rangle_{AD} \left| P \bar{P} \right\rangle_{BC}$ &$\left| \Phi^+ \right\rangle_{AD} = \frac{1}{\sqrt{2}} \left(\left| P \right\rangle_A \left| P \right\rangle_D + \left| \bar{P} \right\rangle_A \left| \bar{P} \right\rangle_D \right)$  \\
\hline\hline
\end{tabular}
\end{table}

The above discussion shows that if one wish to transfer a C-symmetrical packaged entangled states $\left| \Psi^\pm \right\rangle_{AB}$, then he/she needs to choose the identical combinations, i.e., $\left| \Psi^\pm \right\rangle_{AB} \left| \Psi^\pm \right\rangle_{CD}$ and $\left| \Phi^\pm \right\rangle_{AB} \left| \Phi^\pm \right\rangle_{CD}$. However, if one wish to transfer a C-asymmetrical packaged entangled states $\left| \Phi^\pm \right\rangle_{AB}$, then he/she needs to choose the cross combinations, i.e., $\left| \Psi^\pm \right\rangle_{AB} \left| \Phi^\pm \right\rangle_{CD}$.

\subsection{Matter-antimatter asymmetry via packaged entangled states}

Matter-antimatter asymmetry (or baryogenesis) \cite{Dine,Morrissey,Weinberg,Liddle} usually refers to the imbalance between baryons and antibaryons produced after the Big Bang.
The observations have shown that our universe has far more matter than antimatter \cite{Canetti,Ahlen,Cohen,Steigman}. 
This means that our universe either started with more matter than antimatter, or started with equal amount of matter and antimatter but later became matter dominated due to some unknown physical laws. 
 The former seems to be in conflict with relativistic quantum mechanics\cite{Dirac} and the latter were usually studied by research works.\cite{Dine}
Up to now, a number of mechanisms have been proposed to address the matter-antimatter asymmetry, such as Higgs field baryogenesis \cite{Kusenko}, electroweak baryogenesis \cite{Kuzmin,Dolgov}, leptogenesis \cite{Fukugita,Davidson}, Affleck-Dine baryogenesis \cite{Affleck}, GUT (grand unified theory) baryogenesis \cite{Weinberg2,Kolb}, Planck-scale baryogenesis \cite{Dine}, etc.

We will now present a new interpretation to the origin of the matter-antimatter asymmetry using the theory of packaged entangled states. Eq.(\ref{PhiE}) shows that the total charges of the packaged entangled states $\left| \Phi^\pm \right\rangle_{AB}$ are not conserved in the process of wave function collapse. This phenomenon directly leads to the imbalance between particles and antiparticles, or particle-antiparticle asymmetry. Therefore, the collapse of packaged entangled states with C-symmetry breaking may contribute to the matter-antimatter asymmetry of our universe.

First, we will propose a process of ``entanglement selection'' to explain why the particles created after the Big Bang are in the packaged entangled states, but not in the separable states.
Second, we will show that strong interaction and electromagnetic interaction resulted in a phase transition and therefore leaded to the baryon/anti-baryon asymmetry. 
Finally, we will show that the collapse of packaged entangled states with C-symmetry breaking satisfies the Sakharov conditions \cite{Sakharov}.
Because the collapse of wave function occurred after particle creation, the second quantization formalism is not necessary and first quantization formalism is sufficient to describe the wave function collapse.

\subsubsection{Entanglement selection}

The matter-antimatter asymmetry can be described by the asymmetry parameter \cite{Komatsu}
$\eta_B = \left(n_B-n_{\bar{B}}\right)/n_\gamma \simeq n_B/n_\gamma = \left(6.19 \pm 0.15\right) \times 10^{-10}$,
where $n_B$, $n_{\bar{B}}$, and $n_\gamma$ are the overall number density of baryons, antibaryons, and cosmic background radiation photons, respectively.
One can see that $\eta_B$ is a very small number. Thus, a mechanism only needs to account for a tiny imbalance produced at a very early time.

\textbf{(1) Quark creation and annihilation}.
It is believed that the asymptotically free quarks, gluons, and leptons were created in Quark Epoch (the first millionth of a second after the Big Bang, i.e., $t<10^{-6}$ s and $kT > 150$ MeV). An equal amount of quarks and anti-quarks (with zero total charge) should be created according to relativistic quantum mechanics \cite{Dirac}. 
This condition is obviously satisfied by both the separable states $\left| \Theta^\pm \right\rangle_{AB}$ in Eq.(\ref{ThetaS}), and the packaged entangled states $\left| \Psi^\pm \right\rangle_{AB}$ in Eq.(\ref{PsiE}) and $\left| \Phi^\pm \right\rangle_{AB}$ in Eq.(\ref{PhiE}). Each state that satisfies the particle-antiparticle pair condition is possible to occur according to the principles of quantum mechanics.

If the quarks and anti-quarks were created in the separable states $\left| \Theta^\pm \right\rangle_{AB}$ in Eq.(\ref{ThetaS}), then there should be equal amount of quarks and anti-quarks. The quarks would annihilate all the anti-quarks.
This is also true for the packaged entangled states $\left| \Psi^\pm \right\rangle_{AB}$ in Eq.(\ref{PsiE}), which have $C$-symmetry and the law of charge conservation holds in the wave function collapse.
More specifically, if $A$ collapse into a quark $\left| P \right\rangle_A$, then $B$ will collapse into an anti-quark $\left| \bar{P} \right\rangle_B$, i.e., $\left| \Psi^\pm \right\rangle_{AB} \rightarrow \left| P \right\rangle_A \left| \bar{P} \right\rangle_B$. If $A$ collapse into an anti-quark $\left| \bar{P} \right\rangle_A$ , then $B$ will collapse into a quark $\left| P \right\rangle_B$, i.e., $\left| \Psi^\pm \right\rangle_{AB} \rightarrow \pm \left| \bar{P} \right\rangle_A\left| P \right\rangle_B$. In other words, an equal amount of quarks and anti-quarks will be produced after the wave function collapse and they will annihilate each other. It should be emphasized that a quark can only annihilate with an anti-quark of the same flavor. Thus, the quarks and anti-quarks created in Eq.(\ref{ThetaS}) and Eq.(\ref{PsiE}) won't live long and will go back to photons.

However, if the quarks and anti-quarks were created in the packaged entangled states $\left| \Phi^\pm \right\rangle_{AB}$ in Eq.(\ref{PhiE}), then the collapse of these wave function will break C-symmetry. The law of charge conservation does not hold in this process.
More specifically, if $A$ collapse into a quark $\left| P \right\rangle_A$, then $B$ will also collapse into a quark $\left| P \right\rangle_B$, i.e., $\left| \Phi^\pm \right\rangle_{AB} \rightarrow \left| P \right\rangle_A\left| P \right\rangle_B $. If $A$ collapse into an anti-quark $\left| \bar{P} \right\rangle_A$, then $B$ will also collapse into an anti-quark $\left| \bar{P} \right\rangle_B$, i.e., $\left| \Phi^\pm \right\rangle_{AB} \rightarrow \pm \left| \bar{P} \right\rangle_A \left| \bar{P} \right\rangle_B $. In other words, an unequal amount of quarks and anti-quarks were produced after the wave function collapse. Thus, the quarks and anti-quarks created in Eq.(\ref{PhiE}) may survive and live long.

\textbf{(2) Phase transition}.
From a collective point of view, however, even if every quark-antiquark pair were created in the packaged entangled states $\left| \Phi^\pm \right\rangle_{AB}$, it is still possible that an equal amount of quarks and anti-quarks will be produced after the collapse of all states $\left| \Phi^\pm \right\rangle_{AB}$. This is because there is an equal probability for each $\left| \Phi^\pm \right\rangle_{AB}$ to collapse into either $\left| P \right\rangle_A \left| P \right\rangle_B$ or $\pm \left| \bar{P} \right\rangle_A \left| \bar{P} \right\rangle_B$. After the collapse of all these $\left| \Phi^\pm \right\rangle_{AB}$, the total number of quarks will be equal to that of anti-quarks. These quarks and anti-quarks will annihilate each other and go back to photons. Fortunately, this did not happen due to the strong interaction between quarks and gluons which leaded to a phase transition. Let us now prove it.

After the packaged entangled states $\left| \Phi^\pm \right\rangle_{AB}$ collapse into quarks (or antiquarks), the strong interaction between the quarks (or antiquarks) will reduce the quarks' (or antiquarks') energy and combine them into baryons (or antibaryons). Generally, the quark-quark potential due to the strong interaction between quarks can be written as $V_{qq} = -\frac{4}{3}\frac{\alpha_s}{r} + kr$, where $\alpha_s$ is the quark-gluon coupling, $r$ is the distance between quarks, and $k$ is confinement constant.\cite{Griffiths,Perkins} Because each baryon is made up of three quarks, the total quark-quark potential of each baryon can be approximately written out as $V_{Baryon} \approx 3 \left( -\frac{4}{3}\frac{\alpha_s}{r_A} + kr_A\right) $, where $r_A$ is the average distance between quarks.

Recall the definition of baryon number (density) $B=\frac{1}{3}\left(n_q-n_{\bar{q}}\right)$, where $n_q$ is the number (density) of quarks and $n_{\bar{q}}$ is the number (density) of antiquarks. According to this definition, a baryon has a baryon number of $+1$ and an antibaryon has a baryon number of $-1$. Thus, the total potential of quarks in the baryons (or antibaryons), $V_t$, can be written out as,
\begin{equation}
V_t \approx 3 \left(-\frac{4}{3}\frac{\alpha_s}{r_A} + kr_A\right) \left|B\right|,
\end{equation}
where $\left|B\right|$ is the absolute value of $B$.

Let us now use symbol $E_0$ to denote energy density of all quarks created in the packaged entangled states $\left| \Phi^\pm \right\rangle_{AB}$ (before the collapse of wave function). If the collapse of all packaged entangled states result in an equal amount of quarks and antiquarks, then these quarks and antiquarks will annihilate each other and go back to photons. The total energy will remain in the value of $E_0$. However, if all the packaged entangled states collapse into quarks (or antiquarks), then the energy density of quarks (or antiquarks) will reduce to $E_0 - V_t$. Collecting all these terms, the total energy density of quarks can be written out as,
\begin{equation}
\label{TotalEnergy}
E_t = \left\{ \begin{array}{ll}
E_0 & \textrm{before the collapse of the packaged entangled states}\\
E_0 - V_t & \textrm{after the collapse of the packaged entangled states}
\end{array} \right.
\end{equation}

Eq.(\ref{TotalEnergy}) shows that the energy density of all quarks, $E_t$, is a decreasing function of the absolute value of baryon number (density) $B$ (see Fig. 3). 
The packaged entangled states are states with higher energy ($E_0$). However, the strong interaction between quarks (or antiquarks) made it energy favorable for more packaged entangled states to collapse into baryons (or antibaryons). They experienced a process of spontaneous symmetry breaking and therefore leaded to a phase transition that resulted in matter-antimatter asymmetry.

\begin{figure}[htb]
	\label{figure3}
	\begin{center}
		\includegraphics[height=0.45\textwidth]{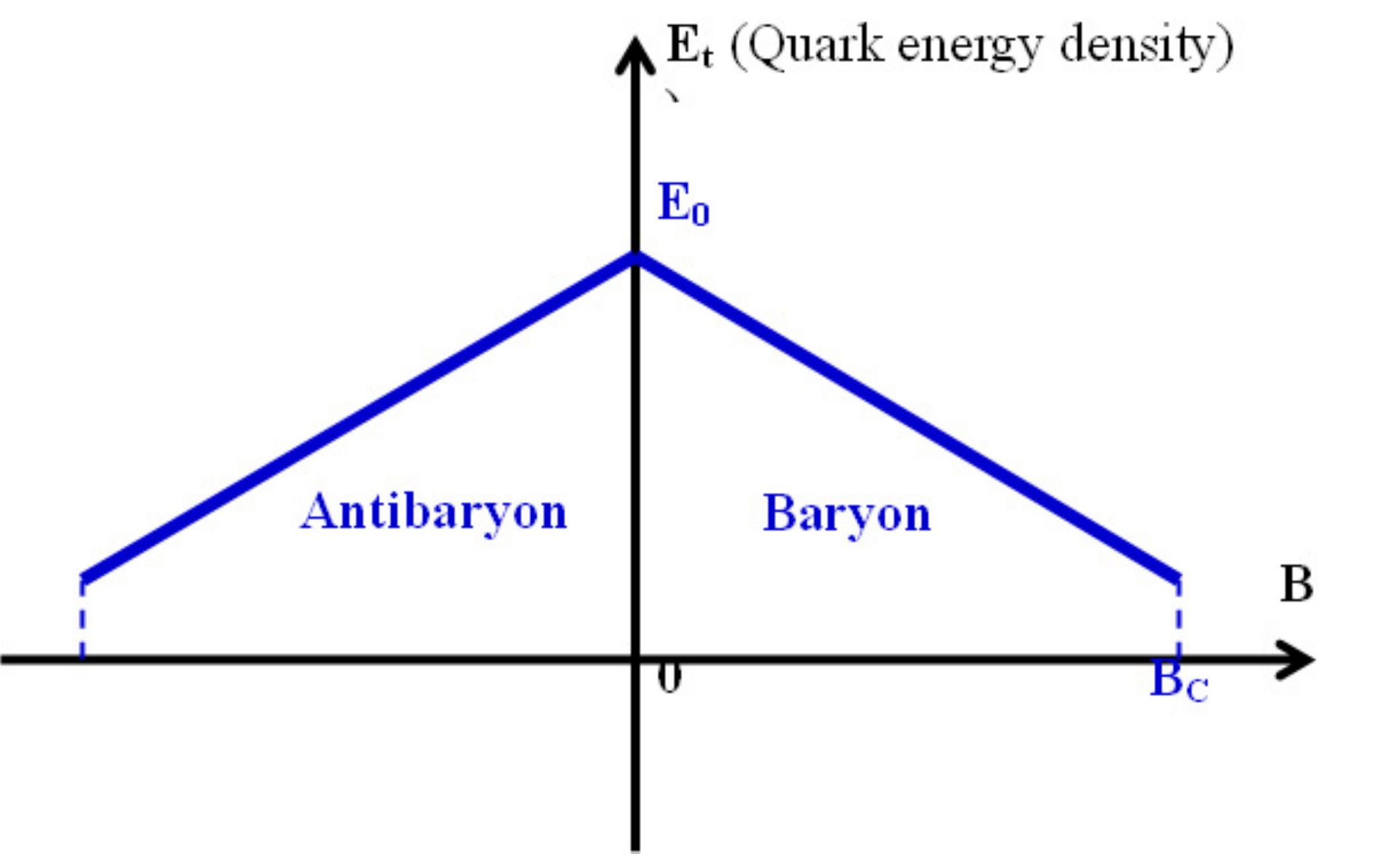}
		\caption{(Color online) Schematic diagram for the quark energy density $E_t$ as a function of baryon number $B$. An spontaneous symmetry breaking occurs due to the quark-quark potential after the collapse of packaged entangled states. With the expansion of universe, the baryon number stop increasing at the value $|B_C|$ where the total energy density (or temperature) of the universe reduced to the lowest value for creating new quarks/anti-quarks ($150$ MeV).
		}
	\end{center}
\end{figure}

With expansion of the universe, the temperature reduce to a critical value where the energy density is not enough to create new quarks/anti-quarks ($kT < 150$ MeV). 
Finally, part of the baryons will annihilate an equal amount of antibaryons as the universe expanded and cooled down. However, the remnant particles that are not annihilated will survive. This should contribute to the tiny residue asymmetry parameter $\eta_B$. \cite{Komatsu}

\textbf{(3) Leptogenesis}.
It should be also mentioned that, once the imbalance between baryons and antibaryons is formed, there is a non-zero net charge which will generate an electric field. This field will interact with the leptons/antileptons in packaged entangled states. The interaction is described by the Coulomb potential $V_{em}=-\frac{\alpha}{r}$. Thus, the collapse of the packaged entangled state of leptons ($\left| \Phi^\pm \right\rangle_{AB}$) is not random, but has a partiality depending on the electric field generated by the baryons (or antibaryons). In other words, the electric field baryons (or antibaryons) will cause the leptons/antileptons in the packaged entangled field to collapse into particles with a charge conjugating to that of the baryons (or antibaryons). This means that the baryogenesis results in leptogenesis.

Now we can summarize the entanglement selection process based on the packaged entangled states with C-symmetry breaking: \textit{due to particle-antiparticle annihilation, the particles created in the separable states and packaged entangled states with C-symmetry are short-lived, but the particles created in the packaged entangled states with C-symmetry breaking can be long-lived and survive until today due to phase transition caused by the quark-quark potential}.
The entanglement selection process is similar to the idea of universal selection from the universal Darwinism (generalized Darwinism) \cite{Dawkins,Smolin,Zurek}.

\subsubsection{Packaged entangled states and Sakharov conditions}

In 1967, Sakharov \cite{Sakharov} proposed three necessary conditions for a physical process that can lead to the matter-antimatter asymmetry: 1. Violation of baryon number B,
2. Violation of C-symmetry and CP-symmetry,
3. Departure from thermal equilibrium.
We shall now choose the packaged entangled states $\left| \Phi^+ \right\rangle_{AB}$ (see Eq.(\ref{PhiP})) to show that the collapse of a packaged entangled state with C-symmetry breaking satisfies the Sakharov conditions \cite{Sakharov}.

\textbf{(1) Violation of baryon number $B$}.
All quarks ($u$, $d$, $t$, $b$, $c$, $s$) carry a baryon number \cite{Griffiths,Perkins} $B=1/3$ and their anti-quarks carry a baryon number $B=-1/3$.
Let us now assume that, in the packaged entangled state $\left| \Phi^+ \right\rangle_{AB}$ (see Eq.(\ref{PhiP})), the $\left| P \right\rangle$s represent quarks and $\left| \bar{P} \right\rangle$s represent anti-quarks.  
Therefore, the total baryon number of $\left| \Phi^+ \right\rangle_{AB}$ is $B=0$. Under the external perturbation, if $\left| \Phi^+ \right\rangle_{AB}$ collapses into the separable state $\left| P \right\rangle_A \left| P \right\rangle_B$, then the variation of baryon number is $\Delta B = 2/3$. However, if $\left| \Phi^+ \right\rangle_{AB}$ collapses into the separable state $\left| \bar{P} \right\rangle_A \left| \bar{P} \right\rangle_B$, then the variation of baryon number is $\Delta B = - 2/3$. This means that the total baryon number $B$ is violated in either case during the collapse of wave function.

Similarly, if the $\left| P \right\rangle$s and $\left| \bar{P} \right\rangle$s in $\left| \Phi^+ \right\rangle_{AB}$ represent the leptons and anti-leptons, then the total lepton number $L$ is violated during the collapse of wave function. In fact, the collapse of $\left| \Phi^+ \right\rangle_{AB}$ also violates the other packaged quantum numbers, i.e., $Q$, $I_3$, $C$, $S$, $T$, and $B'$.

\textbf{(2) Violation of C-symmetry and CP-symmetry}.
Recall that $\left| \Phi^+ \right\rangle_{AB}$ is an eigenstate of the charge conjugation operator $C$ (see Eq.(\ref{OperatorC})). Applying $C$ to $\left| \Phi^+ \right\rangle_{AB}$, we have
\begin{equation}
\label{VioCSym1}
C \left| \Phi^+ \right\rangle_{AB} = \left| \Phi^+ \right\rangle_{AB}.
\end{equation}

For the separable states $\left| P \right\rangle_A \left| P \right\rangle_B$ and $\left| \bar{P} \right\rangle_A \left| \bar{P} \right\rangle_B$, we have
\begin{equation}
\label{VioCSym2}
C \left(\left| P \right\rangle_A \left| P \right\rangle_B \right) = \left| \bar{P} \right\rangle_A \left| \bar{P} \right\rangle_B, ~\text{and}~ C\left(\left| \bar{P} \right\rangle_A \left| \bar{P} \right\rangle_B\right) = \left| P \right\rangle_A \left| P \right\rangle_B.
\end{equation}

The C-symmetry is obviously conserved in Eq.(\ref{VioCSym1}), but not conserved in Eq.(\ref{VioCSym2}). This shows that the C-symmetry breaks in the process of wave function collapse: $\left| \Phi^+ \right\rangle_{AB} \rightarrow \left| P \right\rangle_A \left| P \right\rangle_B$, or $\left| \Phi^+ \right\rangle_{AB} \rightarrow \left| \bar{P} \right\rangle_A \left| \bar{P} \right\rangle_B$.

The parity operator $P$ reverses the space coordinates, i.e., $P: \vec{r} \rightarrow -\vec{r}$. Applying $P$ to $\left| \Phi^+ \right\rangle_{AB}$, we have $P\left| \Phi^+ \right\rangle_{AB} = p \left| \Phi^+ \right\rangle_{AB}$, where $p=\pm 1$ is the eigenvalues of $P$. The exact value of $p$ depends on the exchange symmetry of the particle states. Similarly, we have $P \left(\left| P \right\rangle_A \left| P \right\rangle_B \right) = p \left| P \right\rangle_A \left| P \right\rangle_B $ and $P \left(\left| \bar{P} \right\rangle_A \left| \bar{P} \right\rangle_B \right) = p \left| \bar{P} \right\rangle_A \left| \bar{P} \right\rangle_B $.

Now applying the product operator $CP$ to $\left| \Phi^+ \right\rangle_{AB}$, we have
\[CP \left| \Phi^+ \right\rangle_{AB} = p \left| \Phi^+ \right\rangle_{AB}, \]

Applying $CP$ to the separable states $\left| P \right\rangle_A \left| P \right\rangle_B$ and $\left| \bar{P} \right\rangle_A \left| \bar{P} \right\rangle_B$, we have
\[CP \left(\left| P \right\rangle_A \left| P \right\rangle_B \right) = p \left| \bar{P} \right\rangle_A \left| \bar{P} \right\rangle_B, ~\text{and}~ CP \left(\left| \bar{P} \right\rangle_A \left| \bar{P} \right\rangle_B \right) = p \left| P \right\rangle_A \left| P \right\rangle_B. \]

The product $CP$ cannot send $\left| P \right\rangle_A \left| P \right\rangle_B$ or $\left| \bar{P} \right\rangle_A \left| \bar{P} \right\rangle_B$ back to $\left| \Phi^+ \right\rangle_{AB}$. Thus, the CP-symmetry \cite{Sakurai,Peskin} breaks in the process of wave function collapse: $\left| \Phi^+ \right\rangle_{AB} \rightarrow \left| P \right\rangle_A \left| P \right\rangle_B$ or $\left| \Phi^+ \right\rangle_{AB} \rightarrow \left| \bar{P} \right\rangle_A \left| \bar{P} \right\rangle_B$.

\textbf{(3) Departure from thermal equilibrium}.
In fact, the expansion of universe assures the departure from thermal equilibrium.\cite{Weinberg}. But we would like to show that the collapse of packaged entangled state with C-symmetry breaking further assures the departure from thermal equilibrium.

The particles in packaged entangled state $\left| \Phi^+ \right\rangle_{AB}$ are indeterminate. Now consider that $\left| \Phi^+ \right\rangle_{AB}$ collapses into a separable state, i.e., $\left| \Phi^+ \right\rangle_{AB} \rightarrow \left| P \right\rangle_A \left| P \right\rangle_B (\text{or} \left| \bar{P} \right\rangle_A \left| \bar{P} \right\rangle_B) $. However, the particles in the separable state $\left| P \right\rangle_A \left| P \right\rangle_B$ (or $\left| \bar{P} \right\rangle_A \left| \bar{P} \right\rangle_B$) are determinate, i.e., they are either particles or antiparticles. 
This means that the separable states $\left| P \right\rangle_A \left| P \right\rangle_B$ (or $\left| \bar{P} \right\rangle_A \left| \bar{P} \right\rangle_B$) cannot go back to the packaged entangled state $\left| \Phi^+ \right\rangle_{AB}$.
Therefore, the wave function collapse, $\left| \Phi^+ \right\rangle_{AB} \rightarrow \left| P \right\rangle_A \left| P \right\rangle_B (\text{or} \left| \bar{P} \right\rangle_A \left| \bar{P} \right\rangle_B) $, is an irreversible process. 
Further, the particles in the separable states are identical particles and their cannot annihilate each other. These show that the collapse of packaged entangled state with C-symmetry breaking assures the departure from thermal equilibrium.

\section{Discussion}
\label{Disc}

We have studied the properties of packaged entangled states and their applications. Let us now further discuss the difference by comparing to other theories.

\textit{(1) Characteristics of packaged entangled states}.
The packaged entangled states is strongly related to the charge conjugation operator $C$ which packages a number of quantum numbers ($Q$, $B$, $L$, $I_3$, $C$, $S$, $T$, $B'$) as an entirety. The quantum numbers cannot be added in or taken out separately. The particles in the packaged entangled states are indeterminate. This means that each particle in the packaged entangled states is partially a particle and partially an antiparticle. In other words, a packaged entangled state is an entangled state of particle and antiparticle, i.e., an entangled state of different particles. Apparently, the hyperentanglement and multimode entangled states do not have these properties. 
They are the entangled states of several physical quantities of identical particles. 
These difference result in the difference between the particle-antiparticle teleportation protocol described in this paper and other quantum teleportation protocols described in early literatures \cite{Bennett,Dur}, and the difference between the entanglement transfer described in this paper and quantum swapping described in early literatures.

On the other hand, the charge of a particle and an antiparticle are equal in quantity but with opposite signs, i.e., a particle and an antiparticle are symmetrical in the sense of charge. As discussed before, the particles in a packaged entangled state are a superposition of a particle and an antiparticle. Thus, their charges are indeterminate. If we refer a particle's charge as its gender, then the particles in the packaged entangled states are hermaphroditic particles. In this sense, the identities of the particles in a packaged entangled states are indeterminate.

\textit{(2) Particle-antiparticle teleportation}. 
In early quantum teleportation protocols \cite{Bennett,Dur}, Alice sends out the information of particle $X$ by performing a Bell measurement on particles $X$ and $A$. This measurement has four possible results. Thereafter, Alice needs a classical channel to inform Bob about her Bell measurement result. In the present particle-antiparticle teleportation protocol, however, Alice sent out the information of $X$ by annihilating $X$ with $A$. Alice's experimental result is fixed by the particle-antiparticle annihilation phenomenon. Thus, Bob's result has a fixed relationship with that of $X$ and he can decode the information directly. The classical channel between Alice and Bob is then removed, which indicates the possibility of superluminal communication. This is consistent with the recent experimental results which have shown that the speed of ``spooky action at a distance'' is at least 10,000 times of the speed of light.\cite{Yin}

In the particle-antiparticle teleportation process, Alice can modulate the receiver's particle (particle $B$) to be a particle or an antiparticle at a distance by choosing particle $X$ to be a particle or an antiparticle. Therefore, Alice can control particle $B$ whether to annihilate or not to annihilate with its environment particles at will. 
This property may be applied in medicine, such as positron emission tomography (PET) \cite{Saha}, positron annihilation spectroscopy (PAS), and the band structure measurements in solid state physics as well.
Furthermore, it may also be applied in remote control.
Finally, the particle-antiparticle teleportation process could be used to transport energy because the particle-antiparticle annihilation process can release a large amount of energy.

\textit{(3) Transfer of packaged entangled states}.
In the transfer of packaged entangled states, the outcome is fixed due to the particle-antiparticle annihilation phenomenon. But in quantum swapping \cite{Zukowski,Schmid}, the outcome is not fixed and there are four possible results due to the Bell measurement.

\textit{(4) External physical quantities}.
As mentioned before, the charge conjugation operator $C$ packages a number of internal quantum numbers ($Q$, $B$, $L$, $I_3$, $C$, $S$, $T$, $B'$), but it does not change the particles' mass, energy, momentum, and spin. 
This means that the packaged entangled states do not include these external physical quantities. The relationship between the internal and external physical quantities will be discussed in a forthcoming paper.

\textit{(5) Matter-antimatter asymmetry}.
The present paper used first quantization formalism (quantum mechanics) to describe the mechanism that caused the matter-antimatter asymmetry. The main idea is the collapse of wave function of the packaged entangled states. The matter-antimatter asymmetry was formed after particle creation.
However, the other theories \cite{Kusenko,Kuzmin,Fukugita,Affleck,Weinberg2} used second quantization formalism (quantum field theory) to describe the mechanism that caused the matter-antimatter asymmetry. The main ideas are the interactions between elementary particles. The matter-antimatter asymmetry was formed in the process of particle creation.

The entanglement selection process indicates that it is a random result that our universe is made of matter, but not anti-matter. More specifically, it depends on the initial condition at the moment of wave function collapse. For example, if the packaged entangled states $\left| \Phi^\pm \right\rangle_{AB}$ collapse and happened to roll into $\left| P \right\rangle_A \left| P \right\rangle_B$, then an universe made of matter is created. However, if $\left| \Phi^\pm \right\rangle_{AB}$ happened to roll into $\pm \left| \bar{P} \right\rangle_A \left| \bar{P} \right\rangle_B$, then an anti-universe made of antimatter is created.

Furthermore, the packaged entangled states and entanglement selection process may indicate the possibility of multiverse \cite{Vilenkin,Linde,Tegmark}. This is because the entanglement selection process results in the random origin of the universe, which means that the structure and composition of a universe is not unique. This multiple origin indicates that the existence of universe may be not unique, but multiple.

The matter-antimatter asymmetry via the collapse of packaged entangled states with C-symmetry breaking occurred after particle creation. It does not conflict with the mechanisms proposed in early literatures, which occurred in the process of particle creation. All these mechanisms may coexist.

\section{Conclusion}

We have shown that particles and antiparticles can form packaged entangled states. These states can be divided into two types, i.e., C-symmetrical and C-asymmetrical. The packaged entangled states are the eigenstates of charge conjugation operator. The species of particles in the packaged entangled states are indeterminate. They are superpositions of a particle and an antiparticle.
We proposed a protocol for teleporting a particle or an antiparticle to a large distance using the packaged entangled states. Different to early studies, the particle-antiparticle teleportation protocol introduced here does not need a classical channel due to the particle-antiparticle annihilation phenomenon. One can teleport a particle identical to the original particle to the receiver using the packaged entangled states with C-symmetry, but teleport a particle conjugating to the original particle to the receiver using the packaged entangled states with C-symmetry breaking.
One can also transfer a packaged entangled state from a particle pair to another particle pair. 
Finally, we show that the collapse of packaged entangled states with C-symmetry breaking contributes to the matter-antimatter asymmetry of the observed universe. The particles created in separable states and packaged entangled states with C-symmetry are short-lived, but the particles created in the packaged entangled states with C-symmetry breaking can be long-lived and survive until today. The collapse of packaged entangled states with C-symmetry breaking satisfies the Sakharov conditions, i.e., violation of baryon number B, violation of C-symmetry and CP-symmetry, and departure from thermal equilibrium.


\begin{thebibliography}{0}
	
	
	\bibitem {Horodecki}
	R. Horodecki, P. Horodecki, M. Horodecki and K. Horodecki. Quantum entanglement. Rev. Mod. Phys. 81, 865-942 (2009). DOI:https://doi.org/10.1103/RevModPhys.81.865
	
	
	\bibitem {Lo}
	Hoi-Kwong Lo, S. Popescu and T. Spiller (Editors). Introduction to Quantum Computation and Information (World Scientific, River-Edge, 1998).
	
	
	\bibitem {Nielsen}
	Michael A. Nielsen \& Isaac L. Chuang. Quantum Computation and Quantum Information (Cambridge University Press, New York, 2010).
	
	
	\bibitem {Einstein}
	A. Einstein, B. Podolsky, and N. Rosen. Can Quantum-Mechanical Description of Physical Reality Be Considered Complete? Phys. Rev. 47, 777 (1935). DOI:https://doi.org/10.1103/PhysRev.47.777
	
	
	\bibitem {Einstein2}
	Letter from Einstein to Max Born, 3 March 1947, p158. The Born-Einstein Letters; Correspondence between Albert Einstein and Max and Hedwig Born from 1916 to 1955 (Macmillan Press Ltd., New York, 1971).
	
	
	\bibitem {Bell}
	John S. Bell. On the Einstein-Podolsky-Rosen paradox. Physics 1, 195-200 (1964).
	
	
	\bibitem {Freedman}
	Stuart J. Freedman and John F. Clauser. Experimental Test of Local Hidden-Variable Theories. Phys. Rev. Lett. 28, 938 (1972). DOI:https://doi.org/10.1103/PhysRevLett.28.938
	
	
	\bibitem {Aspect}
	Alain Aspect, Philippe Grangier, and Gérard Roger. Experimental Realization of Einstein-Podolsky-Rosen-Bohm Gedankenexperiment: A New Violation of Bell's Inequalities. Phys. Rev. Lett. 49, 91 (1982). DOI:https://doi.org/10.1103/PhysRevLett.49.91
	
	
	\bibitem {Bennett}
	C. H. Bennett, G. Brassard, C. Crepeau, R. Jozsa, A. Peres, W. K. Wootters. Teleporting an unknown quantum state via dual classical and Einstein-Podolsky-Rosen channels. Phys. Rev. Lett. 70, 1895-1899 (1993). DOI:https://doi.org/10.1103/PhysRevLett.70.1895
	
	
	\bibitem {Bouwmeester}
	Dik Bouwmeester, Jian-Wei Pan, Klaus Mattle, Manfred Eibl, Harald Weinfurter \& Anton Zeilinger. Experimental quantum teleportation. Nature 390, 575-579 (1997). doi:10.1038/37539
	
	
	\bibitem {Boschi}
	D. Boschi, S. Branca, F. DeMartini, L. Hardy, \& S. Popescu. Experimental Realization of Teleporting an Unknown Pure Quantum State via Dual Classical and Einstein-Podolsky-Rosen Channels. Phys. Rev. Lett. 80, 1121-1125 (1998). DOI:https://doi.org/10.1103/PhysRevLett.80.1121
	
	
	\bibitem {Leuenberger}
	Michael N. Leuenberger, Michael E. Flatte, and D. D. Awschalom. Teleportation of Electronic Many-Qubit States Encoded in the Electron Spin of Quantum Dots via Single Photons. Phys. Rev. Lett. 94, 107401 (2005). DOI:https://doi.org/10.1103/PhysRevLett.94.107401
	
	
	\bibitem {Pfaff}
	W. Pfaff, B. J. Hensen, H. Bernien, S. B. van Dam, M. S. Blok, T. H. Taminiau, M. J. Tiggelman, R. N. Schouten, M. Markham, D. J. Twitchen, R. Hanson. Unconditional quantum teleportation between distant solid-state quantum bits. Science 345, 532 (2014). DOI: 10.1126/science.1253512
	
	
	\bibitem {Krauter}
	H. Krauter, D. Salart, C. A. Muschik, J. M. Petersen, Heng Shen, T. Fernholz \& E. S. Polzik. Deterministic quantum teleportation between distant atomic objects. Nature Physics 9, 400-404 (2013). doi:10.1038/nphys2631
	
	
	\bibitem {Hofmann}
	Julian Hofmann, Michael Krug, Norbert Ortegel, Lea Gerard, Markus Weber, Wenjamin Rosenfeld, Harald Weinfurter. Heralded Entanglement Between Widely Separated Atoms. Science 337, 72-75 (2012). DOI: 10.1126/science.1221856
	
	
	\bibitem {Riebe}
	M. Riebe, H. Haffner, C. F. Roos, W. Hänsel, J. Benhelm, G. P. T. Lancaster, T. W. Karber, C. Becher, F. Schmidt-Kaler, D. F. V. James \& R. Blatt. Deterministic quantum teleportation with atoms. Nature 429, 734-737 (2004). doi:10.1038/nature02570
	
	
	\bibitem {Barrett}
	M. D. Barrett, J. Chiaverini, T. Schaetz, J. Britton, W. M. Itano, J. D. Jost, E. Knill, C. Langer, D. Leibfried, R. Ozeri, D. J. Wineland. Deterministic quantum teleportation of atomic qubits. Nature 429, 737-739 (2004). doi:10.1038/nature02608
	
	
	\bibitem {Olmschenk}
	S. Olmschenk, D. N. Matsukevich, P. Maunz, D. Hayes, L.-M. Duan, C. Monroe. Quantum Teleportation Between Distant Matter Qubits. Science 323, 486-489 (2009). DOI: 10.1126/science.1167209
	
	
	\bibitem {Nakamura}
	Y. Nakamura, Yu. A. Pashkin, and J. S. Tsai. Coherent control of macroscopic quantum states in a single-Cooper-pair box. Nature 398, 786-788 (1999). doi:10.1038/19718
	
	
	\bibitem {Baur}
	M. Baur, A. Fedorov, L. Steffen, S. Filipp, M. P. da Silva, and A. Wallraff. Benchmarking a Quantum Teleportation Protocol in Superconducting Circuits Using Tomography and an Entanglement Witness. Phys. Rev. Lett. 108, 040502 ( 2012). DOI:https://doi.org/10.1103/PhysRevLett.108.040502
	
	
	\bibitem {Kwiat}
	P. G. Kwiat. Hyper-entangled states. J. Mod. Opt. 44, 2173 (1997). doi: 10.1080/09500349708231877
	
	
	\bibitem {Barreiro}
	Julio T. Barreiro, Nathan K. Langford, Nicholas A. Peters, and Paul G. Kwiat. Generation of Hyperentangled Photon Pairs. Phys. Rev. Lett. 95, 260501 (2005). DOI:https://doi.org/10.1103/PhysRevLett.95.260501
	
	
	\bibitem {Chen}
	Jun Chen, Jingyun Fan, Matthew D. Eisaman, and Alan Migdall. Generation of high-flux hyperentangled photon pairs using a microstructure-fiber Sagnac interferometer. Phys. Rev. A 77, 053812 (2008). DOI:https://doi.org/10.1103/PhysRevA.77.053812
	
	
	\bibitem {Vallone}
	Giuseppe Vallone, Raino Ceccarelli, Francesco De Martini, and Paolo Mataloni. Hyperentanglement of two photons in three degrees of freedom. Phys. Rev. A 79, 030301(R) (2009). DOI:https://doi.org/10.1103/PhysRevA.79.030301
	
	
	\bibitem {Liu}
	Kui Liu, Jun Guo, Chunxiao Cai, Shuaifeng Guo, and Jiangrui Gao. Experimental Generation of Continuous-Variable Hyperentanglement in an Optical Parametric Oscillator. Phys. Rev. Lett. 113, 170501 (2014). DOI:https://doi.org/10.1103/PhysRevLett.113.170501
	
	
	\bibitem {Gatti}
	A. Gatti, R. Zambrini, M. San Miguel, and L. A. Lugiato. Multiphoton multimode polarization entanglement in parametric down-conversion. Phys Rev A 68, 053807 (2003). DOI:https://doi.org/10.1103/PhysRevA.68.053807
	
	
	\bibitem {Giovannetti}
	Vittorio Giovannetti, Diego Frustaglia, and Fabio Taddei. Electronic Hong-Ou-Mandel interferometer for multimode entanglement detection. Phys Rev B 74, 115315 (2006). DOI:https://doi.org/10.1103/PhysRevB.74.115315
	
	
	\bibitem {Tan}
	Hua-tang Tan,Wen-wu Deng and He Huang. Multimode entanglement and squeezing from coupled four-wave mixing oscillators. J. Phys. B: At. Mol. Opt. Phys. 43, 215507 (2010). DOI:https://doi.org/10.1088/0953-4075/43/21/215507
	
	
	\bibitem {Shi}
	Wenxing Shi, Fei Wang, Lihui Zhang, Zhiming Zhan, Xing Li. Continuous-variable multimode entanglement in multi-wave mixing. Optics Communications 285, 4446–4452 (2012). DOI:http://dx.doi.org/10.1016/j.optcom.2012.06.047
	
	
	\bibitem {Liew}
	T C H Liew and V Savona. Multimode entanglement in coupled cavity arrays. New Journal of Physics 15, 025015 (2013). DOI:https://doi.org/10.1088/1367-2630/15/2/025015
	
	
	\bibitem {Knott}
	P. A. Knott, T. J. Proctor, Kae Nemoto, J. A. Dunningham, and W. J. Munro. Effect of multimode entanglement on lossy optical quantum metrology. Phys Rev A 90, 033846 (2014). DOI:https://doi.org/10.1103/PhysRevA.90.033846
	
		
	\bibitem {Marinatto}
	Luca Marinatto, Tullio Weber. Teleportation with indistinguishable particles. Phys. Lett. A 287, 1 (2001). DOI:http://dx.doi.org/10.1016/S0375-9601(01)00431-5
	
	
	\bibitem {Kulik}
	S. P. Kulik, S. N. Molotkov, and S. S. Straupe. On teleportation in a system of identical particles. JETP Letters 92(3), 188 (2010). doi:10.1134/S0021364010150130
	
	
	\bibitem {Marzolino}
	Ugo Marzolino and Andreas Buchleitner. Performances and robustness of quantum teleportation with identical particles. Proc. R. Soc. A 472, 20150621 (2016). DOI:https://doi.org/10.1098/rspa.2015.0621
		
	
	\bibitem {Griffiths}
	D. J. Griffiths. Introduction to Elementary Particles (Wiley-VCH, 2nd ed., 2008).
	
	
	\bibitem {Perkins}	
	Donald H. Perkins. Introduction to High Energy Physics (Cambridge University Press, 4th Edition, 2000).	
		
	
	\bibitem {Krauss}
	Lawrence M. Krauss. The Physics of Star Trek (Flamingo, Reissue edition, 1995).
	
	
	\bibitem {Dine}
	Michael Dine and Alexander Kusenko. Origin of the matter-antimatter asymmetry. Rev. Mod. Phys. 76, 1 (2003). DOI:https://doi.org/10.1103/RevModPhys.76.1
	
	
	\bibitem {Peskin}
	M. E. Peskin, D. V. Schroeder. An introduction to quantum field theory (Addison-Wesley, 1995).	
	
	
	\bibitem {BoundPair}
	Not all particle (antiparticle) states, but only neutral systems (with zero total charge) are the eigenstates of the charge conjugation operator $C$,\cite{Griffiths} such as $\gamma$, $\pi^0$, and a bound ``particle-antiparticle'' pair $\left|P\bar{P}\right\rangle$ etc. The bound pair $\left|P\bar{P}\right\rangle$ is an eigenstate of the charge conjugation operator $C$ because it satisfies the relation \cite{Griffiths} $C (\left|P\bar{P}\right\rangle) =  (-1)^J\left|P\bar{P}\right\rangle$, where $J=L+S$ is the total angular momentum quantum number, $L$ is the orbital angular momentum quantum number, and $S$ is the total spin quantum number. For bosons $\left|b\right\rangle$, $S$ is an even number and the relation reduces to $C (\left|b\right\rangle \left|\bar{b}\right\rangle) = (-1)^L \left|b\right\rangle \left|\bar{b}\right\rangle$. 
	
	
	\bibitem {Aharonov}
	Yakir Aharonov and Leonard Susskind. Charge Superselection Rule. Phys. Rev. 155, 1428 (1967). DOI:https://doi.org/10.1103/PhysRev.155.1428
	
	
	\bibitem {Rolnick}
	William B. Rolnick. Does Charge Obey a Superselection Rule? Phys. Rev. Lett. 19, 717 (1967). DOI:https://doi.org/10.1103/PhysRevLett.19.717
	
		
	\bibitem {Hermitian}
	Apply the charge conjugation operator $C$ to the particle state $\left|P\right\rangle$ twice, we have $C^2\left|P\right\rangle= C \left|\bar{P}\right\rangle= \left|P\right\rangle$. This gives $C^2=1$. On the other hand, the normalization condition $ \left\langle P \left| C^{\dag} C \right| P \right\rangle = \left\langle \bar{P} | \bar{P} \right\rangle = 1 $ gives $C^\dag C =1$. Thus, we have $C=C^\dag$.
	
	
	\bibitem {Leinaas}
	Jon Magne Leinaas, Jan Myrheim, and Eirik Ovrum. Geometrical aspects of entanglement. Phys. Rev. A 74, 012313 (2006). DOI:https://doi.org/10.1103/PhysRevA.74.012313
	
	
	\bibitem {Anni}
	When a low-energy elementary particle-antiparticle pair annihilates, say an electron annihilates with a positron, they can only produce two or more photons. But when a composite particle-antiparticle pair annihilates, say a proton annihilates with an antiproton, they will produce multiple particles, such as photons, electrons, positrons, and neutrinos.
	
	
	\bibitem {Klempt}
	Eberhard Klempt, Chris Batty, Jean-Marc Richard. The antinucleon–nucleon interaction at low energy: Annihilation dynamics. Physics Reports 413 (4–5), 197–317 (2005). DOI:http://dx.doi.org/10.1016/j.physrep.2005.03.002
	
	
	\bibitem {Schlosshauer}
	M. Schlosshauer. Decoherence, the measurement problem, and interpretations of quantum mechanics. Rev. Mod. Phys. 76, 1267–1305 (2005). DOI:https://doi.org/10.1103/RevModPhys.76.1267
	
		
	
	
	\bibitem {Zwillinger}
	D Zwillinger. CRC Standard Mathematical Tables and Formulae. (32 edition, CRC Press, 2011)
	
	
	\bibitem {GHZ}
	Daniel M. Greenberger, Michael A. Horne, Anton Zeilinger. Going Beyond Bell's Theorem. arXiv:0712.0921 (2007).
	https://arxiv.org/abs/0712.0921
	
	
	
	\bibitem {Zhao}
	Zhi Zhao, Yu-Ao Chen, An-Ning Zhang, Tao Yang, Hans J. Briegel \& Jian-Wei Pan. Experimental demonstration of five-photon entanglement and open-destination teleportation. Nature 430 54-58 (2004).
	doi:10.1038/nature02643
	
	
	\bibitem {Dur}
	W. D\"{u}r \& J. I. Cirac. Multiparty teleportation. Journal of Modern Optics 47 (2-3), 247-255 (2000). DOI:http://dx.doi.org/10.1080/09500340008244039
	
	
	\bibitem {Zukowski}
	M. Zukowski, A. Zeilinger, M. A. Horne, and A. K. Ekert. ‘‘Event-ready-detectors’’ Bell experiment via entanglement swapping. Phys. Rev. Lett. 71, 4287 (1993). DOI:https://doi.org/10.1103/PhysRevLett.71.4287
	
	
	\bibitem {Schmid}
	Christian Schmid, Nikolai Kiesel, Ulrich K Weber, Rupert Ursin, Anton Zeilinger and Harald Weinfurter. Quantum teleportation and entanglement swapping with linear optics logic gates. New Journal of Physics 11, 033008 (2009). DOI:https://doi.org/10.1088/0031-8949/11/3/033008
	
	
	
	
			
	\bibitem {Morrissey}
	David E Morrissey and Michael J Ramsey-Musolf. Electroweak baryogenesis. New Journal of Physics 14, 125003 (2012). DOI:https://doi.org/10.1088/1367-2630/14/12/125003
	
	
	\bibitem {Weinberg}
	Steven Weinberg. Cosmology (Oxford University Press, New York, 2008).
	
	
	\bibitem {Liddle}
	Andrew Liddle. An Introduction to Modern Cosmology (Wiley, West Sussex, England, 3rd Edition, 2015).
	
	
	\bibitem {Canetti}
	Laurent Canetti, Marco Drewes and Mikhail Shaposhnikov. Matter and antimatter in the universe. New Journal of Physics 14, 095012 (2012). DOI:https://doi.org/10.1088/1367-2630/14/9/095012
	
	
	\bibitem {Ahlen}
	S. P. Ahlen, S. Barwick, J. J. Beatty, C. R. Bower, G. Gerbier, R. M. Heinz, D. Lowder, S. McKee, S. Mufson, J. A. Musser, P. B. Price, M. H. Salamon, G. Tarle, A. Tomasch, and B. Zhou. New Limit on the Low-Energy Antiproton/Proton Ratio in the Galactic Cosmic Radiation. Phys. Rev. Lett. 61, 145 (1988). DOI:https://doi.org/10.1103/PhysRevLett.61.145
	
	
	\bibitem {Cohen}
	A. G. Cohen, A. De Rujula, and S. L. Glashow. A Matter-Antimatter Universe? Astrophys. J. 495, 539 (1998). DOI:10.1086/305328
	
	
	\bibitem {Steigman}
	G. Steigman. The Radio Continuum Morphology of Spiral Galaxies. Annu. Rev. Astron. Astrophys. 14, 336 (1976). DOI:10.1146/annurev.aa.14.090176.002221
	
	
	\bibitem {Dirac}
	P. A. M. Dirac. The Quantum Theory of the Electron. Proc. R. Soc. Lond. A 117 (778), 610-624 (1928). DOI:10.1098/rspa.1928.0023; The Quantum Theory of the Electron. Part II. Proc. R. Soc. Lond. A 118 (779), 351-361 (1928). DOI:10.1098/rspa.1928.0056; A Theory of Electrons and Protons. Proc. R. Soc. Lond. A 126 (801), 360-365 (1930). DOI:10.1098/rspa.1930.0013
	
	
	
	\bibitem {Kusenko}
	Alexander Kusenko, Lauren Pearce, and Louis Yang. Postinflationary Higgs Relaxation and the Origin of Matter-Antimatter Asymmetry. Phys. Rev. Lett. 114, 061302 (2015). DOI:https://doi.org/10.1103/PhysRevLett.114.061302
	
	
	\bibitem {Kuzmin}
	V. Kuzmin, V. Rubakov and M. Shaposhnikov. On anomalous electroweak baryon-number non-conservation in the early universe. Phys. Lett. B 155, 36 (1985). DOI:http://dx.doi.org/10.1016/0370-2693(85)91028-7
	
	
	\bibitem {Dolgov}
	A. Dolgov. Non-GUT baryogenesis. Phys. Rep. 222, 309 (1992). DOI:http://dx.doi.org/10.1016/0370-1573(92)90107-B
	
	
	\bibitem {Fukugita}
	M. Fukugita and T. Yanagida. Barygenesis without grand unification. Phys. Lett. B 174, 45 (1986). http://dx.doi.org/10.1016/0370-2693(86)91126-3
	
	
	\bibitem {Davidson}
	Sacha Davidson, Enrico Nardi, Yosef Nir. Phys. Rept. 466, 105-177 (2008). Phys. Rept. 466, 105-177 (2008). DOI:http://dx.doi.org/10.1016/j.physrep.2008.06.002
	
	
	\bibitem {Affleck}
	I. Affleck and M. Dine. A new mechanism for baryogenesis. Nucl. Phys. B 249, 361 (1985). DOI:http://dx.doi.org/10.1016/0550-3213(85)90021-5
	
	
	\bibitem {Weinberg2}
	S. Weinberg. Cosmological Production of Baryons. Phys. Rev. Lett. 42, 850 (1979). DOI:https://doi.org/10.1103/PhysRevLett.42.850
	
	
	\bibitem {Kolb}
	E. W. Kolb, and M. S. Turner. The Early Universe (Adddison-Wesley, Reading, MA, 1990).
	
	
	\bibitem {Sakharov}
	A. D. Sakharov. Violation of CP invariance, C asymmetry, and baryon asymmetry of the universe. JETP Lett. 5, 24–27 (1967); republished as A. D. Sakharov, Soviet Physics Uspekhi 34 (5), 392–393 (1991). DOI:https://doi.org/10.1070/PU1991v034n05ABEH002497
		
	
	\bibitem {Komatsu}
	E. Komatsu et al. Seven-Year Wilkinson Microwave Anisotropy Probe (WMAP) Observations: Cosmological Interpretation. Astrophys. J. Suppl. 192, 18 (2011). DOI:https://doi.org/10.1088/0067-0049/192/2/18
	
	

	
	
	\bibitem {Dawkins}
	R. Dawkins. Universal Darwinism. In: Evolution from molecules to man, ed. D. S. Bendall (Cambridge University Press, 1 edition, 1983).
	
	
	\bibitem {Smolin}
	Lee Smolin. The Life of the Cosmos (Oxford University Press, 1997).
	
	
	\bibitem {Zurek}
	Wojciech Hubert Zurek. Quantum Darwinism. Nature Physics 5, 181-188 (2009). doi:10.1038/nphys1202
	
		
	\bibitem {Sakurai}
	J. J. Sakurai, Jim Napolitano. Modern Quantum Mechanics (Addison-Wesley, Second Edition, San Francisco, 2011)
	
	
	
	
	
	\bibitem {Yin}
	Juan Yin, Yuan Cao, Hai-Lin Yong, Ji-Gang Ren, Hao Liang, Sheng-Kai Liao, Fei Zhou, Chang Liu, Yu-Ping Wu, Ge-Sheng Pan, Qiang Zhang, Cheng-Zhi Peng, Jian-Wei Pan. Lower Bound on the Speed of Nonlocal Correlations without Locality and Measurement Choice Loopholes. Phys. Rev. Lett. 110, 260407 (2013). DOI:https://doi.org/10.1103/PhysRevLett.110.260407
	
	
	\bibitem {Saha}
	Gopal B. Saha. Basics of PET Imaging: Physics, Chemistry, and Regulations (Springer, 2nd ed., 2010)
	
		
	\bibitem {Vilenkin}
	Alexander Vilenkin. Birth of inflationary universes. Phys. Rev. D 27, 2848 (1983). DOI:https://doi.org/10.1103/PhysRevD.27.2848
	
	
	\bibitem {Linde}
	A. D. Linde. Eternally existing self-reproducing chaotic inflanationary universe. Physics Letters B 175 (4), 395–400 (1986). DOI:http://dx.doi.org/10.1016/0370-2693(86)90611-8
	
	
	\bibitem {Tegmark}
	Max Tegmark. Parallel Universes. Scientific American 288 (5), 40-51 (2003). doi:10.1038/scientificamerican0503-40
	
	
			
\end{thebibliography}
\end{document}